\documentclass[11pt,preprint]{aastex}
\pdfoutput=1
\usepackage{graphicx}

\let\oldfootsep=\footnotesep
\def\VEV#1{\left\langle #1\right\rangle}
\newcommand\ltsima{$\; \buildrel <\over\sim \;$}
\newcommand\simlt{\lower.5ex\hbox{\ltsima}}
\newcommand\gtsima{$\; \buildrel >\over\sim \;$}
\newcommand\simgt{\lower.5ex\hbox{\gtsima}}

\newcommand\mearth {{M_\oplus}}

\newcommand\pac{Paczy{\'n}ski }
\newcommand\half{{1\over 2}}
\newcommand\ie{{i.e.}}

\shorttitle{}
\shortauthors{Bennett}

\begin{document}


\title{An Efficient Method for Modeling High-Magnification\\
         Planetary Microlensing Events}

\author{David P. Bennett}
\affil{University of Notre Dame\\
         Department of Physics\\
         Notre Dame, IN 46556, USA}
\email{bennett@nd.edu}


\begin{abstract}
I present a previously unpublished method for calculating and
modeling multiple lens microlensing events that is based
on the image centered ray shooting approach of \citet{em_planet}. It
has been used to model all a wide variety of binary and triple lens systems, but it
is designed to efficiently model
high-magnification planetary microlensing events, because these
high-magnification events are, by far, the most challenging events to
model. It is designed 
to be efficient enough to handle complicated microlensing events, which
include more than two lens masses and lens orbital motion.
This method uses a polar coordinate
integration grid with a smaller grid spacing in the radial direction than in
the angular direction, and it employs an integration scheme specifically
designed to handle limb darkened sources. I present tests that show
that these features achieve second order accuracy for the light curves
of a number of high-magnification planetary events. They
improve the precision of the calculations by a factor of $>100$
compared to first order integration schemes with the same grid spacing
in both directions (for a fixed number of grid points). This method also 
includes a $\chi^2$ minimization method, based on the 
Metropolis algorithm, that
allows the jump function to vary in a way that allows 
quick convergence to $\chi^2$ minima. Finally, I introduce a global
parameter space search strategy
that allows a blind search of parameter space for light curve models
without requiring $\chi^2$ minimization over a large grid of 
fixed parameters. Instead, the parameter space is explored on a
grid of initial conditions for a set of $\chi^2$ minimizations using the
full parameter space. While this method may be somewhat faster
than methods that find the $\chi^2$ minima over a large grid of parameters,
I argue that the main strength of this method is for events with
the signals of multiple planets, where a much higher dimensional parameter space 
must be explored to find the correct light curve model.

\end{abstract}



\clearpage


\section{Introduction}
\label{sec-intro}
Gravitational microlensing has opened a new window on the study
of extrasolar planets as it is the only method that is currently able to
detect low-mass planets in orbits beyond $1\,$AU. In fact, six of
the ten published microlensing planet discoveries have been of
planets of less than Saturn's mass 
\citep{ogle169,gaudi-ogle109,bennett-ogle109,sumi-ogle368,moa310} 
and two of these have masses below
$10\mearth$ \citep{ogle390,bennett-moa192}. The range of planetary
separations probed by microlensing is particularly relevant to tests
of the core accretion model for planet formation, as microlensing is
particularly sensitive to planets just beyond the ``snow-line" 
\citep{ida_lin,kennedy-searth}
where core accretion predicts that the most massive planets should form.

Seven of these ten published planetary microlensing events
\citep{ogle71,ogle169,gaudi-ogle109,bennett-moa192,
dong-moa400,moa310} have been found in high-magnification
events, which although rare, have a much higher planet
detection probability \citep{griest_saf}. A corollary of this 
is that high-magnification events have significant sensitivity to
events with signals from multiple planets \citep{gaudi_nab_sack} 
and to planets in stellar binary systems.
This point was
demonstrated with the discovery of the Jupiter-Saturn analog
system OGLE-2006-BLG-109Lb,c \citep{gaudi-ogle109,bennett-ogle109}.
Furthermore, this event also demonstrated that microlensing can detect
the orbital motion of a planet when the caustic structures
are sufficiently large as can occur for a massive planet
\citep{dong_ogle71} or a ``resonant" caustic. (When a planet
is close to the Einstein ring, the planetary and central caustics
merge to form a ``resonant" caustic.)

This sensitivity to orbital motion and to systems with more
than two lens masses makes the modeling of these
microlensing events significantly more challenging than other
events. In fact, there is currently a backlog of planetary
microlensing events that appear to have three or more lens masses.
There are three such events discovered through the end of the
2008 Galactic bulge observing season, but the analysis is
complete for only one of these \citep{gaudi-ogle109,bennett-ogle109}.
In contrast, the analysis is complete for 8 of the 9 planetary
microlensing events discovered through the end of 
2008 that can be modeled with a single planet and host star.

In this paper, we present a general method for light curve modeling that is
designed to be able to model the most complicated 
and difficult high-magnification microlensing events. This method has gradually
evolved from the first general method for calculating finite source light curves
for binary lens events \citep{em_planet}. Versions of this modeling method have
been used to analyze possible planetary events observed in the 1990's
\citep{rhie_ben96,bennett-macho_planets,mps-98blg35}, 
to make the first successful real-time predictions of caustic crossings
toward the Galactic bulge and Magellanic Clouds
\citep{iauc96blg3,macho-98smc1}, and to analyze binary lensing events
seen by the MACHO collaboration \citep{macho-lmc9,macho-binaries}.
More recent versions of this code have been used to model all of the
known planetary microlensing events, and this code has played a 
major role in modeling 7 of the 10 published planetary microlensing signals
\citep{bond-moa53,ogle390,ogle169,gaudi-ogle109,bennett-moa192,
bennett-ogle109,sumi-ogle368}. 

This method has
several unique features. In Section~\ref{sec-int}, I present
a numerical integration scheme  with two new features 
that improve its precision by more than
a factor of 100 (for a fixed number of integration grid points).
The first feature is an integration scheme that is specifically designed to
handle the singular derivative at the limb of a limb darkened source
profile. This method also features a polar coordinate integration grid with a much larger
grid spacing in the angular than in the radial direction to take advantage
of the lensing distortion of the images. In Section~\ref{sec-lc_calc},
I present tests of this
method using a number of previously analyzed planetary 
microlensing events. These tests confirm the dramatic improvement
in precision for high-magnification events.

Next, in Section~\ref{sec-markov}, I present an adaptive $\chi^2$
minimization recipe based on the \citet{metrop} algorithm
that is designed to rapidly descend to a minimum of 
a complicated $\chi^2$ surface. This method, along with the
integration scheme described in Section~\ref{sec-int}, was critical
for the analysis of the one double-planet microlensing event
that has been successfully modeled \citep{gaudi-ogle109,bennett-ogle109}.
This event was unusually time consuming to model, due to 
the important effect of the orbital motion of one of the planets, but the 
optimizations described in Section~\ref{sec-int} and \ref{sec-markov} made
the modeling of this event tractable.

A global fit strategy, designed to find all the competitive $\chi^2$ minima
for a microlensing event is presented in Section~\ref{sec-global}, and in 
Section~\ref{sec-global-ex}, I present examples of this method for a number of
single-planet events. Finally, in Section~\ref{sec-conclude}, I discuss ways in 
which this method can be improved and reach some conclusions.

\section{Calculation of Planetary Microlensing Light Curves}
\label{sec-int}

Gravitational lensing by stars and planets can be approximated to
extremely high accuracy by the lens equation for point-masses,
\begin{equation}
w = z - \sum_i {\epsilon_i \over \bar{z} - \bar{x}_i} \ ,
\label{eq-mult_lens}
\end{equation}
where $w$ and $z$ are the complex positions of the source and
image, respectively, and $x_i$ are the complex positions of the 
lens masses. This equation uses dimensionless coordinates, normalized
to the Einstein ring radius of the total lens system mass.
The individual lens masses are represented by
$\epsilon_i$, which is the mass fraction of the $i$th lens mass,
so that $\sum_i \epsilon_i = 1$.

If we assume a point source, then we can derive a formula for the
lensing magnification from the Jacobian 
determinant of the lens equation (and its complex conjugate):
\begin{equation}
J = {\partial w\over \partial z}  {\partial \bar{w}\over \partial \bar{z}} 
  -  {\partial w\over \partial \bar{z}} {\partial \bar{w}\over \partial  z}
  = 1 - \left| {\partial w\over \partial \bar{z}} \right|^2 \ ,
\label{eq-J}
\end{equation}
where 
\begin{equation}
 {\partial w\over \partial \bar{z}}
   =  \sum_i {\epsilon_i \over (\bar{z} - \bar{x}_i)^2} \ .
\label{eq-partw}
\end{equation}
Because eq.~\ref{eq-J} gives the Jacobian determinant of the 
inverse mapping from the image plane to the source plane,
the magnification of each image is given by
\begin{equation}
 A = {1\over |J|} \ ,
\label{eq-AJ}
\end{equation}
evaluated at the position of each image. In order to use eq.~\ref{eq-AJ} to determine
the magnification, we must solve the lens equation, \ref{eq-mult_lens}. For
a lens with two point masses, eq.~\ref{eq-mult_lens} can be inverted to yield
a fifth order complex polynomial equation \citep{double-lens,witt90}. The image
locations for a given source position are roots of this polynomial, but there are
either three or five solutions that correspond to physical image positions.
These polynomial roots
can be found with efficient numerical methods (e.g. 
\citet{num-rec}), and this provides a very quick calculation of binary microlensing
light curves for point sources \citep{mao-pac}. The triple lens version of
eq.~\ref{eq-mult_lens} can be inverted to yield a tenth order polynomial equation,
which corresponds to 4, 6, 8, or 10 physical image solutions \citep{rhie-3lens}.
In general, the lens equation, eq.~\ref{eq-mult_lens}, for $n>1$ point masses has
a minimum of $n+1$ images and a maximum of $5(n-1)$ images
\citep{rhie_5n-1,fund_alg}.

The earliest investigations of the two-point mass lens system light
curves have used this
point source approximation \citep{double-lens,mao-pac}. And these point-source
solutions are an important aspect of the \citet{em_planet} method for calculating
multiple-lens light curves, which I further develop in this paper.
The triple lens solution
to  eq.~\ref{eq-mult_lens} was critical for the modeling of the first double-planet 
microlensing event \citep{gaudi-ogle109,bennett-ogle109}. However,
for systems with extreme mass ratios, such as lens systems with masses 
similar to the Sun, Earth, and Moon, 
\citep{gest-sim} standard double precision (64-bit)
arithmetic is insufficient to solve the tenth order polynomial. So, it may be
necessary to resort to quadruple precision (128-bit) calculations, which can
be up to 100 times slower than double precision in some compiler implementations.

Extensions of the point-source approximation have been provided by 
\citet{pej_hey} and \citet{gould-hex}, who
have developed a power series corrections to the point-source
approximation. The \citet{gould-hex} analysis goes to out to the hexadecapole term.
This approximation is valid much closer to the caustics of the multiple
lens light curve than the point-source approximation. This can result in a dramatic
improvement in computation time for microlensing events with cusp-approaches,
but no caustic crossings, such as OGLE-2005-BLG-71 \citep{dong_ogle71}. 
Events with a source trajectory that runs parallel to a caustic for a long
period of time can also see a significant improvement. But for most
events with caustic crossings, the expected improvement in computational time
is much more modest - probably not exceeding a factor of two \citep{gould-hex}.

The majority of the computational effort required to compute light curves
for planetary microlensing events is devoted to the numerical integrations
necessary for finite source calculations. The most obvious method would be
to simply integrate the point-source magnification pattern over the disk of
the source star. However, this method presents severe numerical difficulties
due to the singularities in the point-source magnification profile. The point-source
magnification pattern for a source crossing a fold caustic is 
\begin{eqnarray}
A &\approx C_1 + {C_2\over \sqrt{x - x_c}}  \ \ \ \  {\rm for} \ \ x > x_c
 \nonumber \\
    & \approx C_1  \ \ \ \  {\rm for} \ \ x < x_c \ ,
\label{eq-fcaustic}
\end{eqnarray}
where $C_1$ and $C_2$ are constants, and $x_c$ is the location of the fold
caustic (which is assumed to be parallel with the $y$-axis). So, the point-source
magnification is (formally) infinite and discontinuous on the line-like caustic
curve. In the caustic exterior, there are also pole singularities in the magnification,
$A\approx 1/r$, in the vicinity of cusps.

These singularities can be avoided with the ray-shooting method, originally
developed by \citet{ray-shoot1}. For complicated lens systems that consist of
more than just a few point masses, solving for the image positions for a given
source position can be difficult, or even intractable. But, if we start with the
image positions, we can always use eq.~\ref{eq-mult_lens} to determine the
source position from the image positions. By covering the image plane with
light rays that are shot back towards the source it is possible to find all the 
images for a given source position. This is often referred to as the 
inverse ray-shooting method because without a solution to eq.~\ref{eq-mult_lens},
it is only possible to do the lens mapping in the inverse direction: from image to
source.

For binary and triple lens systems, it is straight forward to solve the lens
equation numerically, so it is not necessary to shoot the rays in the inverse direction.
Instead, the advantage of ray shooting is that the integrands involved in the
lens magnification calculations are less singular. The lens images have
a surface brightness equal to the surface brightness of the source, so we
can integrate in the image plane and avoid the strong singularities associated
with caustic and cusp crossings in the source plane.

This leads to the basic strategy developed by \citet{em_planet} for the calculation
of multiple lens light curves. The lens equation is solved to locate the images
for a point-source located at the source center. If these images are sufficiently
far from the critical curves and the source sufficiently far from the caustics, then
the point-source approximation is used. (In most cases, the point source approximation
is used if these separations are greater than the 7 source radii times the point
source magnification.) When the point-source
approximation cannot be used, we build integration grids in the image plane to
cover each of the images to be integrated over. The image grids are increased until
their boundaries are completely comprised of points that do not map onto the source.
Some care in bookkeeping is required to ensure that images are not double counted
as some grids can grow to include more than one image. There will be times when
the source limb has crossed a caustic, while the source center remains on the
exterior. To include these partial lensed images we must also build integration grids
at the critical curve locations corresponding to the caustic points that overlap the
source.

These finite source calculations become quite time consuming for high-magnification
events due to the nature of the high-magnification images. As the lens and source
approach perfect alignment, we approach the Einstein ring situation, and the images
become large circular arcs with a length:thickness ratio that is approximately
equal to the total magnification, $A$, which is typically in the range 
$100 \simlt A \simlt 1000$. As I shall discuss below, it is these long, thin images
that are very time consuming to integrate over, and so many of the features
of my numerical integration scheme are designed to make these integrals
more efficient.

For static lens systems (where the orbital motion of the lens system is not important)
the image positions at different times in the light curve will often overlap, so
we will often have to invoke the lens equation, eq.~\ref{eq-mult_lens}, many times
at the same location. So, to minimize the number of lens equation calculations
for high-magnification events, we store the lens equation solutions on a grid
centered on the Einstein ring. (This feature is common to a number of other
methods that use a version of ray-shooting \citep{wamb,ratt,dong-ogle343}.)

This image centered ray shooting 
light curve calculation strategy \citep{em_planet} was the first method
that was able to give precise calculations of planetary microlensing light curves 
including realistic finite source effects. Several potentially promising alternative 
approaches have been suggested 
\citep{wamb,gould-stokes,ratt,dong-ogle343,dominik-cont}. Several of these
focus on finding solutions with fixed lens mass ratios and relative positions
\citep{wamb,ratt,dong-ogle343} in order to map out $\chi^2$ as a function
of these parameters. This is often used for an initial search for solutions, but it
is not very efficient when these parameters are not fixed. Furthermore, the
number of parameters that must be fixed increases from two to five when 
a triple lens system is considered, and the brute force method of mapping
out the $\chi^2$ surface seems much less attractive if it must be done
in five dimensions. The methods of \citet{gould-stokes}, \citet{dominik-cont},
and the ``loop-linking" method of \citet{dong-ogle343} are somewhat
more flexible. Both \citet{gould-stokes} and \citet{dominik-cont} invoke
a Stokes theorem approach that is much more efficient for sources
that are not limb darkened. Of course, no limb-darkening is an unphysical
approximation, but in many cases, the effect of limb darkening on
microlensing light curves is relatively small. Thus, it might be
sensible to develop a fast code based on the \citet{gould-stokes}
method to search for approximate solutions without limb darkening.
However, the errors due to the lack of limb darkening could be more
serious in events like OGLE-2005-BLG-390 \citep{ogle390}
and MOA-2007-BLG-400 \citep{dong-moa400} where planetary
signal comes from a caustic curve that is much smaller than the source.
The efficiency advantage of the Stokes theorem approach is lost when
limb darkening is included, but this method may still  be competitive.
The basic method of \citet{em_planet} is not tied to specific fixed parameters
and does not provoke these limb-darkening concerns, and therefore
it seems sensible to continue with this basic approach.

\subsection{Integration of Limb Darkened Profiles}
\label{sec-int1}

In order to understand how to write a numerical integration scheme for
gravitational lensing light curves, let us first consider the
simpler question of one-dimensional numerical integration.  There has
been a lot of work in this field, and there are a number of numerical
integration schemes that can give quite precise results for a small number
of integration grid points if the integrand is smooth. A good discussion of these methods
is given in \citet{num-rec}, and here I reproduce the relevant points.

For most numerical integration problems, the key to an efficient evaluation of
the integrals, is to obtain high accuracy with as few evaluations of the integrand
function as possible. This is often accomplished by invoking a higher order
integration scheme, which means that the error can be expected to scale as
a high power of the integration grid spacing, $h$. Of course, high order is
no guarantee of high accuracy. A high order scheme can have a large
coefficient in front of the error term that can render it less accurate than a lower
order scheme at a given grid spacing. Furthermore, in our case, we
are considering two dimensional integrals, so correlations between the numerical
errors in different rows of one dimensional integration can have a significant
effect on the overall accuracy of the integral. That is, there might be correlations
that tend to make the numerical errors in different rows add coherently, instead
of incoherently so that the relative error would fall as the square root of the
number of integration rows. As a result, it is quite difficult to predict the accuracy
of a numerical integration scheme with analytic arguments like the ones 
presented in this section. So, as is usually the case with numerical calculations,
it will be the numerical tests of the method that will show which methods are
most precise.

The
integrands that we are concerned with are not very smooth, so I restrict
the discussion to 2nd order integration schemes. For most numerical
integration problems,
there are two basic building
blocks for the 2nd order integration schemes, the trapezoidal rule, 
\begin{equation}
\int_{x_1}^{x_2} f(x) dx = h\left(\half f_1 + \half f_2 \right) + O(h^3 f^{\prime\prime}) \ ,
\label{eq-trap}
\end{equation}
and the mid-point rule,
\begin{equation}
\int_{x_{1/2}}^{x_{3/2}} f(x) dx = hf_1 + O(h^3 f^{\prime\prime}) \ .
\label{eq-midp}
\end{equation}
These are both formulae for evaluating integrals over a single grid
spacing, $h$, using values of the function calculated at integer
multiples of the grid spacing. The function values are $f_i\equiv f(x_i)$.
The error term $O(\ )$ indicates that the true answer differs from
the estimate by an amount that is the product of some numerical coefficient times
$h^3$ times the value of the second derivative of the function somewhere 
in the range of integration.

Now, these building block formulae can be strung together to 
build extended formulae that can be used over finite intervals. This
yields the extended trapezoidal rule, 
\begin{equation}
\int_{x_1}^{x_N} f(x) dx = h\left(\half f_1 + f_2 + f_3 + ... + f_{N-1} 
    + \half f_N \right) + O\left( {(x_N-x_1)^3 f^{\prime\prime}\over N^2}\right) \ ,
\label{eq-trap_ext}
\end{equation}
and the extended mid-point rule,
\begin{equation}
\int_{x_{1/2}}^{x_{N+1/2}} f(x) dx = h\left( f_{1} + f_{2} + ... + f_{N-1} 
    + f_{N} \right) + O\left( {(x_{N+1/2}-x_{1/2})^3 f^{\prime\prime} \over N^2}\right) \ .
\label{eq-midp_ext}
\end{equation}
Usually, the extended mid-point rule (eq.~\ref{eq-midp_ext})
is presented using integrand values
evaluated at half integer grid points, but we have offset this grid by half a grid 
spacing so that most of the grid points coincide with those of the extended 
trapezoidal rule (eq.~\ref{eq-trap_ext}). When written this way, the 
extended mid-point rule , the extended trapezoidal rule and the new
integration formula presented below will require that the integrand
only be evaluated at integer grid points in the interior. This makes it
clear that the only difference between these integration formulae the
treatment of the boundary.

One might imagine that we could implement something like eq.~\ref{eq-trap_ext}
or \ref{eq-midp_ext} in two dimensions give an integration scheme with a
precision proportional to the inverse square of the total number of grid points.
However, there are two difficulties with this procedure. First is the fact that
our problem is slightly different from the normal numerical integration problem,
because we calculate the $f_i$ values before we know where the boundary is.
Thus, it is impossible for us to arrange that the boundaries are located at 
integer or half-integer values of the grid spacing. If we are interested in
an integration scheme that is only first order accurate, then we only need
to determine which points on the grid are inside the image boundary without
attempting to locate the boundary. However, for a second order method, we
do need to locate the boundary to a precision much greater than the grid spacing.
I solve for the position of the boundary using the Brent's method
\citep{num-rec} to find the boundary to a precision of $0.1h^2$, which $h$
is the grid spacing. This increases the number of lens equation 
(eq.~\ref{eq-mult_lens}) calculations that must be done per row by a factor of
less than 1.5.

The second complication is that our integrands are not very smooth.
Moving the finite source integration from the source plane to the image
plane removes the singularities from the integrand, but for limb darkened
sources, there are still singularities in the derivatives of the surface 
brightness in the image plane, and these will limit our attempt to do these
lens magnification integrals efficiently.

The linear limb darkening law of \citet{milne21} gives a reasonable approximation
to the limb darkening for most stars. This linear law is given by
\begin{equation}
I = I_0 \left[ 1 - c\left(1-\sqrt{1-\rho^2}\right)\right] \ ,
\label{eq-limbd}
\end{equation}
where $I_0$ is the central intensity, $c$ is the linear limb darkening coefficient,
and $\rho$ is the distance from the center toward the limb of the star at $\rho=1$.
This limb darkening law has a first derivative that diverges at $\rho=1$. Also, when
the source crosses a caustic, the fraction of the stellar profile that is inside the
caustic has two additional images that meet on a critical curve in the image
plane. These images have opposite parity, and the first derivative of the surface
brightness profile will have a discontinuity on the critical curve, so the second
derivative will diverge.

We should note that the linear limb darkening model is not a perfect match
to model atmospheres, with an average difference of $> 1\,$\% 
\citep{hey07} from Kurucz's model atmospheres 
\citep{kurucz93a,kurucz93b,kurucz93c,kurucz94}. This discrepancy can be
reduced by going to more complicated limb darkening models
\citep{claret2000}. However, the microlensing
light curves involve integrals over the limb darkened profiles, and these are
generally much more accurate 
than the limb darkened profiles themselves. Furthermore, there is 
no guarantee that these models are actually correct, so it is perhaps
more sensible to compare with previous well sampled microlensing
events that yielded high precision limb darkening measurements.
The first such example is event MACHO 95-BLG-30 \citep{macho-95blg30}
which is the first example of an event detected in progress which exhibited 
finite source effects. They employ a ``quadratic" limb darkening model
in place of eq.~\ref{eq-limbd}, but due to sparse sampling, the 
$\chi^2$ improvement with respect to a model without limb darkening was
only $\Delta\chi^2 = 9$, so it is likely that the improvement over
the linear model is quite small. The high cadence follow-up observations
of the PLANET collaboration yielded a number of binary lens caustic
crossing events with a much stronger limb darkening signal. For example,
PLANET found that limb darkening improved the fit for the
binary microlensing event MACHO-97-BLG-28 \citep{planet-97blg28}
by $\Delta\chi^2 = 393$.
However, the additional improvement with the ``square-root" limb darkening
model was only $\Delta\chi^2 = 5$, which they argue is not statistically
significant. One of the most spectacular binary microlensing events
ever observed was event EROS-2000-BLG-5 \citep{planet-er2000b5},
which had a very extended
caustic crossing with a duration of $\sim 4\,$days followed by a cusp
approach to within $0.1$ source radii from the stellar limb four days
later. These features were measured with hundreds of photometric
measurements while the source was magnified to a brightness ranging 
from $I = 13$ to $I = 15$ using several $1\,$m class telescopes. It is difficult
to imagine circumstances that would allow a higher S/N measurement of
limb darkening effects. However, the non-linear limb darkening parameters
are found to be $< 0.1\sigma$ away from 0. So, this event does not
yield a significant measurement of limb darkening parameters 
beyond the linear term. Similar results were also obtained for
the well sampled single lens event OGLE-2004-BLG-254
\citep{cassan-ogle254}.

It is perhaps somewhat more instructive to consider high-magnification
events. since these are the events that are the focus of the method
presented in this paper. Such events also show only weak evidence that
non-linear limb darkening models improve the fits.
\citet{moa2002blg33} find
marginal evidence for terms beyond the linear term of eq.~\ref{eq-limbd}
for the $V$ band 
in the MOA-2002-BLG-33 light curve, an event with a very strong caustic
crossing at the peak. But, there was no evidence for a term beyond the linear
one in the $I$ band data, which dominate the light curve coverage for
most events. Similarly, \citet{dong-moa400} and \citet{moa310} find only
a modest improvement $\chi^2$ for non-linear limb darkening models
planetary events with geometries that should maximize limb darkening 
effects. More importantly, the more complicated limb darkening models
have no significant influence on the non-limb darkening parameters, so 
there is little reason to consider limb darkening models more complicated
than eq.~\ref{eq-limbd} unless we are specifically interested in the limb darkening
parameters.

For high-magnification events, it is the divergent first derivative at the limb that
is, by far, the most serious problem. The images are highly extended parallel to
the Einstein ring and compressed by a factor of about two in the radial direction,
so they have a very large ratio of boundary to area. Thus, errors at the boundary
make a large contribution to the total error in the integral. The caustic crossing
features, on the other hand, are less singular than the limb darkening profile, and
they also are also (usually) much shorter than the entire extent of the limb in
the image plane.

While eq.~\ref{eq-limbd} describes the limb darkening in the source plane,
the integrals are carried out in the image plane where the source brightness
profile is distorted by the gravitational lens. However, the lowest order
behavior near the limb is generically described by $I \sim C + D\sqrt{x}$
where $C$ and $D$ are constants and $x$ is the distance from the 
limb of the distorted image. The only case where the $\sqrt{x}$ behavior
is removed by the lens distortion is when the stellar limb just touches the
interior of a caustic. However, this will generally only occur at a single
point of contact between the caustic and the limb, so the $\sqrt{x}$ behavior is generic.

The two building block formulae, eqs.~\ref{eq-trap} and \ref{eq-midp}, are
derived by requiring that they be exact for low order power laws (as in
a power series expansion of $f$), and in the second order case, the formulae
are exact for $f = {\rm const.}$ and $f = x$. This fails for limb darkened
profiles, because these cannot be expressed as a power series in $x-x_L$,
where $x_L$ is the location of the limb. Instead, the distorted limb darkening
profile can be expressed as a power series in $\sqrt{x-x_L}$. We can still demand that
the our integration formula is exact for the two leading orders in the power series 
expansion of the integrand. 
In eqs.~\ref{eq-trap} and \ref{eq-midp}, the first term in the power series that does
not vanish scales as $h^3$ (under the assumption that $f$ can be expanded in a 
power series in $x$). 
However, with half-integer powers of $h$ in addition
to integer powers, there are more terms in the power series expansion. 
As a result, the first non-vanishing term in eqs.~\ref{eq-trap} and \ref{eq-midp}
scales as $h^{3/2}$ instead of $h^3$ when $f(x)$ has a limb darkened form
like eq.~\ref{eq-limbd}. In order to
cancel this $h^{3/2}$ error term, we will demand that
our integration formula be exact for  $f = {\rm const.}$ and $f = \sqrt{x-x_L}$, where
$x_L$ is the location of the limb. This requirements lead us to replace 
eq.~\ref{eq-midp} by 
\begin{equation}
\int_{x_L}^{x_{3/2}} f(x) dx = h\left(\half+\delta\right)\left[(1-b)f_L + bf_1\right] \ ,
\label{eq-intlimb}
\end{equation}
where 
\begin{equation}
b = {2\over 3} \sqrt{\delta + \half \over \delta} \ ,
\label{eq-intlimbb}
\end{equation}
and $\delta = (x_1-x_L)/h$. The $\delta$ in the numerator of eq.~\ref{eq-intlimbb} is
somewhat worrisome because $\delta$ can become very small if the
limb happens to come very close to a grid point. Conceivably, this could
lead to a situation, where the error grows very large, even if it is formally of
high order. Therefore, we introduce
another parameter, $\delta_c$, such that eq.~\ref{eq-intlimb} is only invoked
for $\delta \geq \delta_c$. 
When $\delta < \delta_c$, we invoke a standard ``second order" method
that will be converted to 1.5 order by the singular derivative at the 
limb. Any combination of $cf_L + df_1$ will satisfy this criteria as long as
$c + d = 1/2 + \delta$. Experimentation with different $c$ and $d$ values
indicates that $c = \delta_i/3$ and $d = 2\delta_i/3 + 1/2$ is a good choice, so
it is used below.
We will investigate the effect of this $\delta_c$ 
parameter in Section~\ref{sec-lc_calc}.
We can now write an extended numerical integration
rule to take the place of eq.~\ref{eq-midp_ext},
\begin{equation}
\int_{x_{L1}}^{x_{L2}} f(x) dx = h\left( A_1 f_{L1} + B_1 f_{1} + f_{2} + ... + f_{N-1} 
    + B_2 f_{N} + A_2 f_{L2} \right) \ ,
\label{eq-int_rule}
\end{equation}
where $L1$ and $L2$ refer to the the stellar limbs at each limit of the $x$ coordinate
integral, and the $A_i$ and $B_i$ coefficients are given by
\begin{eqnarray}
A_i = \left(\half + \delta_i \right)\left(1-b_i \right) \Theta (\delta_i-\delta_c)
    + {\delta_i \over 3} \Theta(\delta_c - \delta_i) \ ,  \nonumber  \\
B_i = \left(\half + \delta_i \right) b_i  \Theta (\delta_i-\delta_c)
    +  \left({2\over 3}\delta + \half\right)  \Theta(\delta_c - \delta_i) \ ,
\label{eq-AB}
\end{eqnarray}
where $\Theta$ is the Heavyside step function and $\delta_i$ and $b_i$ refer
to $\delta$ and $b$ for the each of the two stellar limbs (at $i = 1\,$, 2)
on the image being integrated.

If we set $\delta_c = 0$, then eq.~\ref{eq-int_rule} is accurate to second order, even though
eq.~\ref{eq-intlimb} has a non-vanishing $h^2$ error term. The reason for this is because
we only invoke eq.~\ref{eq-int_rule} at the limbs, and we use eq.~\ref{eq-midp} for all the 
interior points. Since the limb darkened profile does have a power series expansion in $x$
away from the limb, the error for  eq.~\ref{eq-midp} does scale as $h^3$ in the interior (except
on a critical curve, where it has a $h^{5/2}$ contribution). Thus, it is only the $h^3$ error terms
that get a $1/h$ contribution from the sum. So, formally, eq.~\ref{eq-int_rule} is second order
accurate (with $\delta_c = 0$), while eqs.~\ref{eq-trap_ext} and \ref{eq-midp_ext} are only
accurate to the three halves order for a limb darkened source. 
Of course, with $\delta_c > 0$, eq.~\ref{eq-int_rule} also
gains an error term that scales as $h^{3/2}$, but as we shall see, in some cases, even
with $\delta_c > 0$, eq.~\ref{eq-int_rule} can yield second order accurate results.
In all cases, eq.~\ref{eq-int_rule} with $\delta_c \sim 0.15$ is substantially
more accurate than the first order or $\delta_c = 1.0$ calculations.

It is possible to derive integration formulae that are more complicated than
eqs~\ref{eq-int_rule} and \ref{eq-AB} that are formally 2nd order accurate without
the problem of any of the coefficients growing unreasonably large for any position of
the boundary with respect to the grid spacing. However, experimentation with a
number of such integration formulae has not found any such integration scheme
that gives results as accurate as the scheme
represented by eqs~\ref{eq-int_rule} and \ref{eq-AB}.

\subsection{Two-Dimensional Ray Shooting Integration}
\label{sec-int2d}

In Section~\ref{sec-int1} we developed a one-dimensional numerical integration rule,
eq.~\ref{eq-int_rule}, which is designed to improve the accuracy of the
integration of limb darkened source profiles. But, of course, we will need to do
two dimensional integrals to determine microlensing magnifications. 
The integral in the second dimension is not subject to the divergent 
integrand derivative at the boundary, because this is removed by the
integral in the first direction. (Of course, the limb darkening has the
same behavior in both directions. But the integral in the first direction
is roughly proportion to the length of the row being integrated, and this
generally
goes to zero at the boundaries of the integral in the second direction.)
But, we still must deal with the arbitrary
location of the image boundary. I employ the following 
second order accurate formula for this integration
\begin{equation}
\int_{y_L}^{y_{5/2}} F(y)dy = h\left[\left({3\over 8}+\eta + {\eta^2\over 2}\right)F_1
     + \left({9\over 8}- {\eta^2\over 2}\right)F_2\right] \ ,
\label{eq-inty}
\end{equation}
where $\eta = (y_1-y_L)/h$ and $F(y)$ refers to the integral over the $x$
direction, which has a $y$ dependence that is not made explicit in
eq.~\ref{eq-int_rule}.

With eqs.~\ref{eq-int_rule} and \ref{eq-inty} to handle the numerical integrations,
we can now consider the coordinate system to use for the integrations. In this
context, it is useful to consider the image geometry
for high-magnification microlensing events. Consider a typical high-magnification
event with a magnification of $A = 200$. If there are no companion planets or
stars, then there will be two lensed images. The major image will have the shape
of a circular arc with a magnification of $A_{\rm maj} = 100.5$, and it will be located
just outside the Einstein ring. The minor image will be just inside the Einstein ring on
the opposite side of the lens from the major image, and its magnification will be
$A_{\rm min} = 99.5$. Each image will be compressed in the radial direction by
about a factor of two, so the images will have the form of long, skinny arcs with
a length-to-width ratio of about 200. Thus, the limb darkening profile will vary
200 times more rapidly in the radial direction than in the angular direction.
This strong distortion of the images suggests that a polar coordinate grid
is most appropriate for our problem, and it seems likely that we will require
a much larger grid spacing in the angular direction than in the radial direction.
In fact, the 200:1 distortion of the images for our example would seem to suggest
that a 200:1 grid spacing ratio might be appropriate.
 
However, we must also consider the effect of the planetary lenses that are
the primary motivation for observing high-magnification microlensing events.
The planetary lenses will distort the single lens images, and if there are caustic
crossings, new images will be produced that will not follow the Einstein ring 
as closely as the images that are not significantly influenced by the planet. 
So, during the planetary deviations, this image stretching in the angular
direction may not be quite as severe as in our example. However, it is this
image stretching in the angular direction that is responsible for the high
magnification of these events, so we should expect that the optimal 
integration grid should include some extension in the angular direction.
 
Following the discussion above, I have arrived at the following 
two-dimentional integration strategy.  An integration grid is set up
in polar coordinates ($r$,$\theta$) with a larger grid spacing in the angular than in
the radial direction. In Section~\ref{sec-lc_calc}, we study the effects of
varying this axis ratio. The integration is done using using eq.~\ref{eq-int_rule}
in the radial direction with a fixed value of $\delta_c$. The integrand is given
by the value of the limb darkening profile at the integration point, times $r$,
to give the proper polar coordinate area element. However, the integration
is done in the image plane, while the limb darkening is known in the 
source plane. Thus, we must apply the lens equation, eq.~\ref{eq-mult_lens},
to determine the appropriate source plane point and the limb darkened 
surface brightness that corresponds to the integration grid point in the
image plane. The dependence 
of the light curve calculation precision on the grid size, 
the angular vs.~radial grid spacing ratio, and on $\delta_c$ is investigated in
Section~\ref{sec-lc_calc}.

\subsection{Locating and Building the Integration Grids}
\label{sec-grid}

Since we don't know the extend of the images when we start
to calculate their magnification, we require a scheme to build
an integration grid that covers each image, or at least each
image that requires a finite source calculation. 
It most efficient to have integration grids that don't extend far 
beyond the images because rays are
shot from the image plane to the source plane at every grid point. 

The method of \citet{em_planet} was to build a rectangular 
grid centered on each point source image, and to add rows and
columns to each integration grid until the rows and columns at
the boundaries of each do not contain any grid points inside an
image. With Cartesian coordinates, this method is quite inefficient
for high-magnification events, because the grid must cover
almost the entire Einstein ring disk to integrate over thin
images arcs that spread out over much of the Einstein ring. 
Polar coordinates are much more efficient in this case. Cartesian
coordinates may be more efficient for low-magnification events.
But the light curves of these events are less time consuming, so
computational efficiency is less important. The complication of 
using different grid geometries for different events does not seem
justified by the very minor improvement in efficiency for low
magnification events with a Cartesian grid.

A somewhat more efficient method is to build the grid row-by-row,
with each row extending just as far as the image does, an approach
first implemented by \citet{vermaak00}. The grid
is then extended row-by-row until we have a grid that is surrounded
by a boundary of grid points that are outside of the image. Our tests
have shown that this method is typically about a factor of two faster
than building grids that are ``rectangular" in polar coordinates,
and it is this method that is used for our timing results presented in
Section~\ref{sec-global-ex}.
In both cases, one must check that no images are double-counted.

In addition to building integration grids around at the position of
the point source images (when the point-source approximation
cannot be used), we must also ensure that images associated
with caustic crossing are included when the center of the
source is outside the caustic. For static lens systems, this is
most efficiently done by simply calculating the caustic curve
location and building a grid at the location of any caustic point
that is not included in another integration grid. For lens systems
with orbital motion, this method can become inefficient because the
caustics move and must be recalculated at every time step. Another
method that can also be used is to calculate the number of images
for source points on the boundary of the source. 
Then, a grid can be built at the location of any new image near the
image boundary.

Finally, for static lens systems it is possible to speed up the 
calculations by storing the source positions corresponding to
image positions in an annulus centered on the Einstein ring.
This avoids the need to recalculate the lens equation,
eq.~\ref{eq-mult_lens}, for the same image points for the
magnification integration at different points on the light curve.
This same optimization method is used more extensively
in inverse ray-shooting \citep{wamb} and magnification
map \citep{ratt,dong-ogle343,dong-moa400} methods.

\subsection{What Light Curve Calculation Precision Is Necessary?}
\label{sec-what_acc}

In Section~\ref{sec-lc_calc}, I will present the results of light curve calculation tests
with different values of the grid spacing and different calculation parameters, but
first, it will be helpful to consider how much light of light curve precision is needed
for practical modeling calculations. In Section~\ref{sec-int1}, we discussed the
difference between the simple linear limb darkening models and more complicated
models that do a better job of reproducing the limb darkening seen in 
stellar atmosphere calculations as well as a few microlensing events which have
good light curve coverage at limb crossings. However, even these improved models
have deviations from the stellar models that are a factor of a few smaller than the
deviation from the linear model. Thus, we might still expect light curve errors at the
level of $\simlt 0.1$\% at the limb crossings if these improved limb darkening models, are 
used, compared to errors of perhaps $\sim 0.3$\% 
at the limb crossing with the linear model.
These errors of $\simlt 0.1$-0.3\% are comparable 
to the level of systematic photometry errors
that we expect in the microlensing light curves. These systematic errors are expected to 
affect the entire light curve, instead of just the limb crossings, although they are
also likely to have large correlations that may allow some light curve features, such
as weak caustic crossings, to be measured with a precision of $\sim 0.1$\% 
\citep{ogle169}.

These arguments might suggest that there is no need to calculate the light curves to
a precision better than 0.1\%, but in fact, it is usually the case that higher precision is
needed. The reason for this is that the \citet{metrop} algorithm that is generally 
used for modeling multiple lens light curves  is able to optimize the numerical 
errors, so that they tend to minimize $\chi^2$. If the light curves were calculated
perfectly, the $\chi^2$ surface should usually be smooth over small distances
in parameter space, and the main difficulty in finding the $\chi^2$ minima is to
follow the steep and twisting valleys in $\chi^2$ to the local minimum. However,
the numerical errors act to roughen the $\chi^2$ surface on extremely small
scales. If the numerical calculation errors are similar to the photometric
error bars at the light curve peak, then RMS variation in $\chi^2$ would 
be similar to the number of data points at the peak, which varies between
events, but can often be 50 or more. But with parameters chosen by a
$\chi^2$ minimization scheme, we might expect variations several times larger
than this. Because of this, \citet{dong-ogle343}
advocate that the numerical precision of the light curve calculations be less
than one third of the size of the
error bars. However, numerical errors that are this large
can still cause some difficulty. Since the modeling code tends to select parameters
that allow the numerical error to minimize $\chi^2$, the variation in the
$\chi^2$ seen during a modeling run tends to be much larger than the
RMS value. Therefore, I recommend that random
numerical errors be kept at
$\simlt 10^{-4}$ or at least ten times smaller than the smallest
photometric error bars to avoid difficulties in locating local $\chi^2$
minima due to the roughness of the $\chi^2$ surface.

\section{Light Curve Calculation Tests}
\label{sec-lc_calc}
\begin{figure}
\plotone{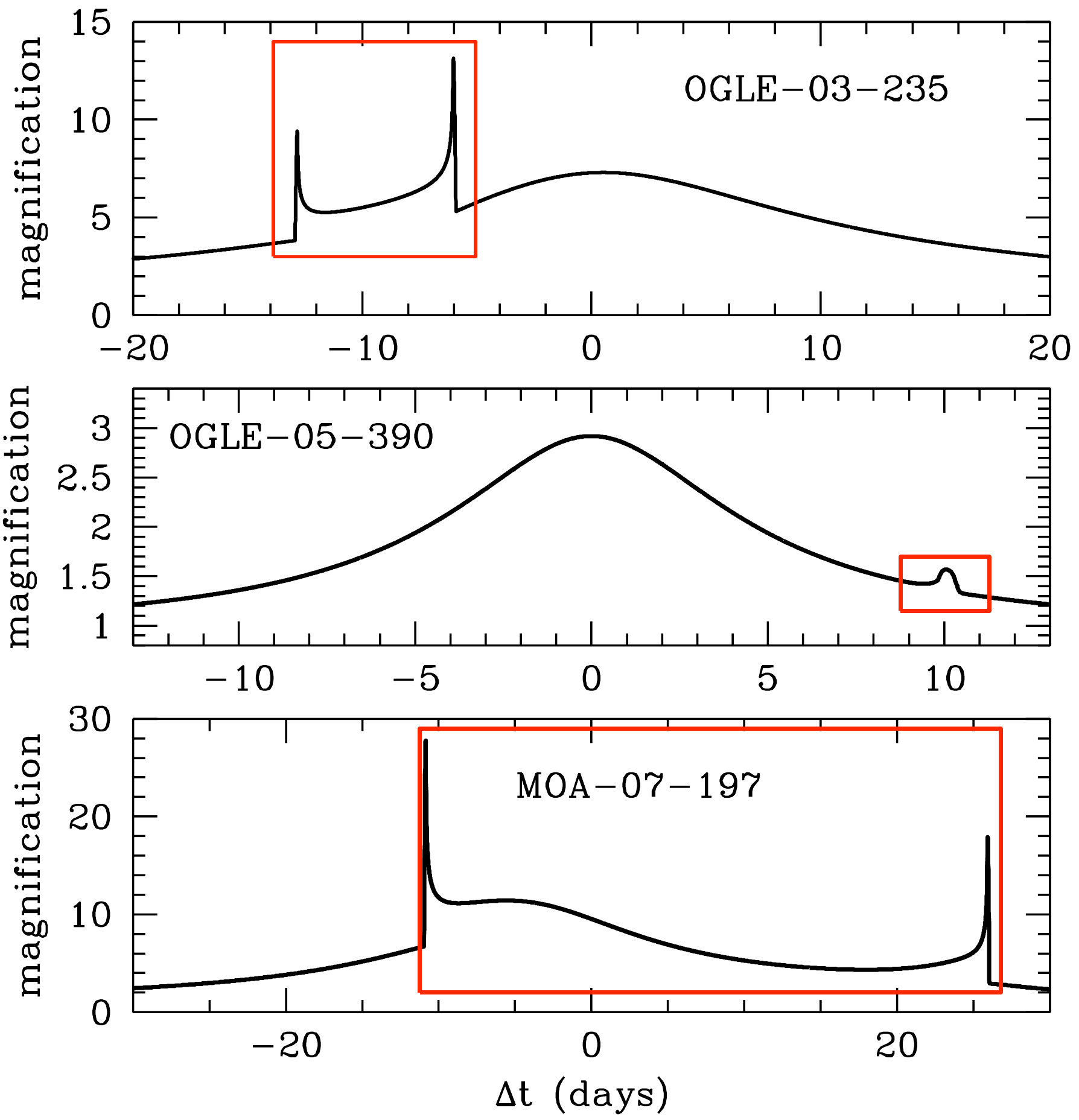}
\caption{Light curves of the low magnification events used in the tests of
calculation precision are shown. Each light curve is a model of an 
observed event with a planetary or brown dwarf secondary lens. The
red boxes indicate the portions of the light curves used for the comparisons.
\label{fig-lc_lo}}
\end{figure}

In order to determine the optimum light curve calculation parameters, I
compare light curves seven different sets of model parameters, 
which are based on models of observed events. Three of these
are relatively low magnification events, which are shown in Figure~\ref{fig-lc_lo}.
These are OGLE-2003-BLG-235, the first definitive planetary microlensing
event \citep{bond-moa53}, OGLE-2005-BLG-390 with a planet of 
$\sim 5\mearth$ \citep{ogle390}, and MOA-2007-BLG-197, which has
a brown dwarf secondary (Cassan, et al. in preparation). 

The other four comparison events are high-magnification events, shown in
Figure~\ref{fig-lc_hi}. These include both cusp approach and caustic 
crossing models for MOA-2007-BLG-192 \citep{bennett-moa192}, which 
includes a planet of $\sim 3\,\mearth$, OGLE-2005-BLG-169 \citep{ogle169},
and MOA-2008-BLG-310 \citep{moa310}. In both of these figures, the red
boxes indicate the regions of the light curves used in tests of light curve
calculation precision. It is important only to use regions of the light curve
where finite source calculations are done. Otherwise, the comparison
of different integration parameters will be diluted by regions where the 
point source approximation is used (and the light curves are identical).

Note that these light curve calculation tests do not always use the
published version of the data set for each event, so the resulting
parameters will sometimes differ slightly from the published ones.
In every case, the values published in the discovery or follow-up analysis
papers should be considered definitive.

Figure~\ref{fig-compare} shows a comparison of the RMS 
fractional precision,
$\sigma$,  as a function of the geometric mean grid spacing,
in units of the source star radius,
for the 3 low-magnification events in the top two panels
and the 4 high-magnification events in the bottom four panels. 
(The geometric mean is the square root of the product of the
angular and radial grid spacings.)
The short-dashed curves have $\delta_c = 1.00$, so that treatment of the
limb-darkening profile is avoided, and the long-dashed black curve is
is a first order integration with no attempt to locate the image boundaries
on sub-grid-spacing scales.
The blue, green, and red curves
have $\delta_c = 0.017$, 0.05, and 0.15, respectively. All the curves
use a angular grid spacing 4 times larger than the radial spacing,
except for the black-dashed curves, which uses equal grid spacings.
(In discussions of the grid spacing, we refer to the ratio for the grids
at the Einstein ring radius. For the high-magnification events, this is
very nearly the exact ratio, since the images are quite close to the
Einstein ring, but for some low magnification events, like 
OGLE-05-390, the planetary deviation images are well outside
the Einstein ring, so the actual grid spacing ratio is somewhat 
larger.) Note that for the low magnification events, the blue and
green curves are often hidden under the red curve.

These comparisons are done with respect to calculations using
$\delta_c = 0.15$ and
an extremely fine grid, with 800 grid points per source radius in 
both the radial and angular directions. I use the RMS fractional deviation,
$\sigma$, between these very high resolution calculations and
the test calculations as the measure of the calculation precision.
The maximum deviation is generally between 2 and $4\times\sigma$,
so there appears to be no significant non-Gaussian tail in the
error distribution.

The curves in Figure~\ref{fig-compare} mostly have a similar slope,
which is close to the
$h^{3/2}$ slope that was predicted in Section~\ref{sec-int1}. However,
for the high-magnification events with $\delta_c = 0.15$, the 
RMS precision scales as $\sigma\sim h^2$. This may seem slightly
surprising because it is only in the $\delta_c \rightarrow 0$
limit, where eqs.~\ref{eq-int_rule} and \ref{eq-AB} achieve second
order accuracy. However, this analysis only applies to a single 
one-dimensional integral, and a full treatment of the accuracy of
the two-dimensional integral must include a number of complications,
such as correlations in the error terms for integrations over different
rows of the two-dimensional domain of integration. Also, it is always
possible for the integration accuracy to scale as a higher power
of $h$ than expected, because the coefficient of the leading order 
term could be so small that a sub-leading term will dominate over
the interesting range of $h$ values. Of course, this is more likely in
situations, such as these limb-darkened source integrals, where the
error terms are power laws in $h^{1/2}$ instead of in $h$.
I expect that this is what has happened for the high-magnification
events with $\delta_c = 0.15$, which have a $\sigma \sim h^2$
scaling despite the fact that the arguments
presented in Section~\ref{sec-int1} suggest that the scaling should
be  $\sigma \sim h^{3/2}$. Thus, the arguments
presented in Section~\ref{sec-int1} and \ref{sec-int2d} should be considered
to be only qualitative, and the comparison with much higher resolution
calculations should be considered to be the definitive measure of the 
numerical errors.

The long-dashed black curves indicate that the first order
calculations seem to do as well as, and often better than
the short-dashed black curves, which represent 
an attempt at a second order correction without the
limb darkening terms given in eqs.~\ref{eq-int_rule} and \ref{eq-AB}.
This might seem somewhat surprising, since the analysis in
Section~\ref{sec-int1} indicates that the one-dimensional integrals
without the limb darkening correction should have errors that
scale as $h^{3/2}$, whereas one-dimensional first order
integration methods have errors that scale as $h$.
\citet{dong-ogle343} have also noted a $h^{3/2}$ error scaling 
in calculations with their method, which is also first order. In
Appendix A.3 of this paper, they note the
improvement from the $\sim h$ error term of the first order
one dimensional integral to the $\sim h^{3/2}$ error term 
observed in the two dimensional integrals, and they attribute
this factor of $h^{1/2}$ improvement in the fractional error
to the $1/N^{1/2}$ Poisson decrease expected if the 
errors in each row are uncorrelated (where $N$ is the number of 
rows). This assumption of
uncorrelated errors is plausible for a first order integration
scheme, but it seems unlikely that a higher order scheme
would also achieve this $h^{1/2}$ improvement when going from one to
two dimensions. So, this might explain why the first order
integration scheme has the same $h^{3/2}$ behavior
as many of the attempted second order integration schemes.
The first order schemes also have the minor advantage that they
don't require additional lens equation (eq.~\ref{eq-mult_lens})
calculations to locate the boundaries on a scale smaller than the
grid spacing, which implies a savings of a factor of up to 1.5 in
computation time. 

The numerical errors for the OGLE-05-390 calculation are significantly
larger than for other events, but this is easily explained by
the details of this event. It is the only event we consider that has
a giant source star. In fact, the planetary caustic responsible
for this planet detection \citep{ogle390} has a diameter that is
4-5 times smaller than the radius of the source, so the caustic crossing
regions of the images are sampled more coarsely than the other
events for the same grid size to source radius ratio. 

For the high-magnification events, 
there is also a clear improvement from increasing the ratio
of the angular to radial grid size from 1 to 4. This effect is
demonstrated even more clearly by Figure~\ref{fig-grid_ratio},
which shows the effect of changing the grid size ratio from
1 to 4 to 16. The high-magnification event calculations show
a clear improvement from the larger ratio of grid sizes, whereas
the lower magnification events show the opposite effect
(with the exception of OGLE-03-235).

In Figure~\ref{fig-improve}, we show the factor by which
$\sigma$ is improved compared to the first order
calculations with an angular:radial grid size ratio of 1. In every
case, the second order scheme with $\delta_c = 0.15$ is
the most accurate, although the improvement is modest for the
low magnification events. (Note that the blue and green curves
for the low magnification events are often hidden under the
red curve.) The low magnification events also
benefit from having an angular:radial grid size 
ratio $= 1$. However, there is
a dramatic improvement for the high-magnification 
events. The improvement ranges from a factor of
10 to a factor of nearly 300 at some grid sizes. The calculations
with $\delta_c = 0.15$ and an angular:radial grid size ratio of 16
prove to be the most accurate for the high-magnification events.
These calculations, represented by the cyan curves in
Figure~\ref{fig-improve}, provide a factor of 100-300 improvement
in precision over the first order calculation case.
If we were to try to reproduce the factor of $\sim 100$ improvement 
seen at the mean grid size of 0.1 by 
simply decreasing the grid size of the
$\delta_c = 1$, grid ratio $= 1$ calculations, we would have to drop the
grid size by a factor of 22. But since this is a 2-dimensional
calculation, this means an increase in the 
number of calculations and hence the computing time by
nearly a factor of 500.

So, in summary, it would seem that the new features that we
have outlined have the potential to increase the computational
efficiency of high-magnification event light curve calculations
by a factor of several hundred.

\section{$\chi^2$ Minimization Recipe}
\label{sec-markov}

In Section~\ref{sec-int}, I developed an efficient method for the
calculation of high-magnification planetary
microlensing light curves, but we also require an efficient method
to move through parameter space to find the $\chi^2$ minimum.
Due to sharp light curve features like caustic crossings
and cusp approaches,
the $\chi^2$ surface for microlensing event models is 
not smooth enough to use a method, like
Levenberg-Marquardt, that make use of the assumed 
smoothness of the $\chi^2$ surface. Instead, we must
use a more robust method, similar to the Markov
Chain Monte Carlo (MCMC) \citep{wmap_mcmc}
or the simulated annealing
method \citep{sim_anneal}, which are both based on the 
\citet{metrop} algorithm. The Metropolis algorithm employs
the Boltzmann factor from statistical physics to decide
whether or not to accept the next proposed step through
parameter space. If the next proposed step reduces
$\chi^2$, (\ie\ $\Delta\chi^2 \leq 0$), then it is always accepted,
but if $\Delta\chi^2 > 0$, then it is accepted
with probability, $e^{-\Delta\chi^2/(2T)}$, where the 
parameter $T$ is referred to as the temperature, in
analogy with statistical physics.

Ideally, we would like to have a scheme that can 
find the global $\chi^2$ minimum automatically without having the
specify any particular initial condition. In fact, the simulated
annealing method was developed to find the global minimum
in situations where there are many local minima.
The basic idea is to start the method with a high temperature,
$T$, in order to explore all of parameter space, and then to 
gradually decrease $T$ and allow the system to relax to the
global $\chi^2$ minimum. While this method has been
used to solve a number of difficult problems, it is not
clear that it will work for planetary microlensing events.
For some events, it might end up in a broad local minimum 
that is favored at high $T$ that is separated by a $\chi^2$
barrier from a narrow, but deeper global minimum. Also, it
is unclear how one would design a schedule for modifying
the temperature that would ensure the efficient location of
the global $\chi^2$ for the observed wide variety of 
planetary microlensing events. Therefore, I do
not attempt to use the Metropolis algorithm to find
the global $\chi^2$ minimum.

A very important aspect of the Metropolis algorithm is the
choice of the jump function.  The jump function starts from the current set of
model parameters and selects a new set of model parameters
to be used to calculate $\chi^2$, which will yield a value of 
$e^{-\Delta\chi^2/(2T)}$ that will allow us to determine whether to move
to this next set of model parameters. 

As mentioned above,
microlensing light curves are characterized by very sharp features
due to caustic crossings and cusp approaches, so we expect that
the $\chi^2$ surfaces we encounter will have steep valleys.
For example, for a caustic crossing event, most directions
in parameter space will cause the timing of a caustic crossing
to change, which will induce a large change in $\chi^2$ if the
photometric measurements sample the caustic very well.
But there will also be some directions in parameter space
that will leave the timing of the caustic crossing fixed. These
will induce much smaller changes in $\chi^2$, so these will
be the directions of the valleys in $\chi^2$ space. An efficient
method of locating and exploiting these $\chi^2$ valleys
has been presented by \citet{cmbeasy}. This involves
calculating the correlation matrix,
\begin{equation}
C_{ij} = \VEV{p_i p_j} = {1\over N} \sum_{k=1}^N p_i p_j \ ,
\label{eq-Cij}
\end{equation}
of the last $N$ (accepted) sets of parameters, $\{p_i\}$.
$C_{ij}$ is then diagonalized, and the diagonalized basis
vectors can be considered to be a new set of parameters
that are uncorrelated over these $N$ steps through parameter
space. We then select new parameters at random from the most
recently accepted parameter set with a Gaussian variance
normalized by the elements of the diagonalized covariance
matrix.

Of course, this scheme cannot be used until after $N$
steps have been accepted, so jump function scheme is needed
for the first $N$ steps. For these initial steps, I specify 
initial uncertainty ranges for each parameter, and select 
new parameters with uniform probability within these ranges.

The use of the Metropolis algorithm in the way I have described
is often referred to as a Markov Chain Monte Carlo (MCMC),
with each accepted step considered a link in the chain. These
MCMC runs can be used to estimate the parameter
uncertainties, but this requires that the jump function be fixed
during the MCMC run. But, such a strategy is not efficient when
searching for a $\chi^2$ minimum, because the $\chi^2$ 
valleys often have many twists and turns, so the minimum
is reached much more quickly when the jump function can
be modified frequently.

I have found that the following recipe allows the
Metropolis algorithm to quickly converge to a local $\chi^2$ minimum
for a wide variety of planetary and binary microlensing events.
The initial jump function is used until $N = \max(20,2N_{\rm par})$
steps have been taken, where $N_{\rm par}$ is the number of
non-linear fit parameters. (Note that if we were to calculate $C_{ij}$
with $N < N_{\rm par}$, we would have a singular matrix, since we
would not have enough points to span the $N_{\rm par}$-dimensional
parameter space.)
Then, the parameter correlation matrix, $C_{ij}$,
is calculated and diagonalized. The new parameters are generated
from the most recent step with a Gaussian probability distribution
following the diagonalized parameter correlation matrix. Of course, these new 
parameters must be converted back to the original non-diagonal parameters
for the light curve calculation and $\chi^2$ evaluation. The
parameter correlation matrix, $C_{ij}$, is recalculated and diagonalized
whenever the number of saved steps, $N$, increases by 4, until
$N$ reaches 100. Once $N = 100$, the oldest saved parameter set 
is dropped each time a new one is added, so that the number of 
saved parameter sets to be used for $C_{ij}$ calculations remains
fixed at $N = 100$, but the $C_{ij}$, is still recalculated and diagonalized
every 4th time that a new parameter set is accepted.

This procedure allows the parameter correlation matrix to gradually
adjust to twists and turns in the $\chi^2$ surface, as the modeling
code travels toward the local $\chi^2$ minimum. However, sometimes
this gradual modification of the parameter correlation matrix is 
not sufficient to keep up with the changing $\chi^2$ surface shape,
and so I also have a procedure for modifying $C_{ij}$ more drastically.
If the code attempts 40 consecutive parameter sets without a single 
one being accepted due to an improvement in $\chi^2$ or passing the
Boltzmann probability test, then the oldest $3/8$ of the parameter
sets are dropped, and if the $N \geq \max(20,2N_{\rm par})$
condition still holds, then 
$C_{ij}$ is recalculated, diagonalized and used to select the next
set of parameters. If $N < \max(20,2N_{\rm par})$, then we revert to
the initial procedure of selecting new parameters with uniform
probability within the initially specified uncertainty ranges. Sometimes
the reduction in the number of saved parameter sets, $N$, will not 
be sufficient to allow a new parameter set to be accepted in the next
40 steps, and in these cases, the number of saved parameter sets
is again reduced by a factor of $5/8$. This procedure can even
be invoked four times in a row to drop $N$ from 100 down to 14.

This $\chi^2$ minimization recipe has been extensively tested and
has been shown to be robust and efficient for finding the local $\chi^2$
minima for a wide variety of microlensing events including all published
planetary microlensing events and the four clear planet detections from
the 2009 bulge observing season, as well as the orbiting 
two planet system, OGLE-2006-BLG-109 \citep{gaudi-ogle109,bennett-ogle109}.
Some examples are discussed in Section~\ref{sec-global-ex}.
Note that this procedure of modifying the jump function should not be
used for a Markov chain calculation that might be used to estimate
parameter uncertainties.

\section{Global Fit Strategy}
\label{sec-global}

In addition to a method for calculating planetary microlensing
event light curves, we also need methods for efficiently
moving through parameter space to find the best fit or fits
(as there are sometimes degeneracies). For events with
only two detectable lens masses and no orbital motion, 
this is fairly straight forward, and a number of methods have
been demonstrated to work. For low magnification planetary
events like OGLE-2005-BLG-390 (Beaulieu et al. 2006), it is possible to 
determine most of the parameters approximately by
inspection of the light curve. For some high-magnification
events, such as OGLE-2005-BLG-71 (Udalski et al. 2005), it is 
also possible to determine the parameters approximately by 
inspection, but it is more prudent to do a systematic search
for solutions. Perhaps the best documented method is the
grid search method of \citet{dong-ogle343,dong-moa400}. In this method,
the best fit is found for each point on a three dimensional
grid over the mass ratio, $q$, lens separation, $d$, and
angle, $\theta$, between the source trajectory and the lens
axis. For each point on this grid, the remaining parameters,
are adjusted using a standard fitting algorithm to find the
$\chi^2$ minimum for the fixed values of $q$, $d$, and
$\theta$. This method generally works quite well, although it
can fail in certain instances, such as the case of
MOA-2007-BLG-192, where there is a degeneracy involving
the source star radius parameter, $t_\ast$, which is usually
not one of the grid parameters. However, this problem is
a result of the sparse light curve sampling for this particular
event, and it seems likely that the magnification map method
can be modified to model this event.

A more serious issue with this grid search method is that
it is impractical to scale it up to systems with more parameters.
If we add another lens mass, this adds three new parameters
(the mass ratio and 2-d position of the additional
mass), to the two parameters (the separation, $d$, and mass ratio, $q$)
that are normally held fixed on the grid. 
If all five of these parameters are not held fixed, then the
computational advantage of the inverse ray shooting method
is lost because the same rays cannot be used throughout the
calculation.
But it is probably too computationally expensive to have more
than three grid parameters. In some cases, it is possible to 
find an approximate solution using a simplified model with fewer
parameters, and then to consider perturbations to this simplified model
to find the full solution. This allows the grid search method to be used
sequentially in stages, first to find the simplified model, and then to
search for the perturbation solutions. This general strategy has proven
to be quite useful for features such as microlensing parallax and lens
orbital motion, as these usually produce only small light curve
perturbations. It has also proven to be effective for some triple
lens models. However, this method will not work for all triple lens
events. Similarly, orbital motion also threatens to derail the
computational advantage of this method, although some strategies
to deal with such problems have been suggested 
\citep{gould-hex}. 
Thus, if we want a general method to model complicated lens systems,
the approach I present here seems more promising. 

I have developed the following method, which has been successfully
tested on virtually all of the planetary microlensing events observed to
date. First we identify the parameters which are obviously well
constrained by the light curve. Typically, this would be the Einstein
ring crossing time, $t_E$, the time, $t_0$, and distance, $u_0$, 
of lens-source closest approach,
the impact parameter, and the source radius crossing time,
$t_\ast$. For events with strong caustic crossings, it may be best to
use one or two of the caustic-limb crossing times in place of
$t_0$ and/or $t_E$. This is similar to the approach advocated
by \citet{cassan-meth}, but it is simpler in that only the time
variables $t_0$ and $t_E$ are modified.
We then set up a coarse grid
over the remaining parameters, and evaluate $\chi^2$ for 
all the grid points. The parameters that yield the best few
$\chi^2$ values are then selected as initial conditions for 
fitting using a modified Markov Chain Monte Carlo (MCMC) routine. 
This procedure is repeated with the
next best $\chi^2$ values from the initial grid search
until we find that most of 
the fits are converging to the same final models. If the fits
converge instead to different models with increasingly worse
$\chi^2$ values, then we repeat this procedure with a denser
initial conditions grid.
This procedure still uses a grid for the initial conditions, but the
modeling runs allow all the parameters to vary. Most of the 
computations for this method are done during these full
modeling runs instead of the initial condition calculations.
As a result, the computation
time does not increase so dramatically with the number of
model parameters.

\section{Global Fit Strategy Examples}
\label{sec-global-ex}

In this section, I present several examples that demonstrate how
the initial condition grid search method works. These examples are
intended to show how to find an approximate solution for each event.
These approximate solutions usually do not include all the solutions
related by the well known light curve degeneracies, such as the
$d \leftrightarrow 1/d$ degeneracy for high-magnification events
\citep{dominik99}.
I include one example
that has not been used for our previous light curve calculation tests,
OGLE-2005-BLG-71. The light curve for
this event can be modeled reasonably well without the inclusion of a  finite source 
\citep{ogle71}. Of course, the source must have a finite size, and the
finite source effect is important for the complete analysis, which is able
to determine the star and planet masses \citep{dong_ogle71}.

This section does include the three high-magnification events that 
have been used in the light curve calculation tests, but of the low-magnification
events used for the light curve calculation tests, only OGLE-2003-BLG-235 is 
included. No systematic effort to search for the correct light curve model 
is needed for OGLE-2005-BLG-390, because it is possible to get a very
good estimate of the parameters by inspection. The single lens parameters
are well determined by a single lens fit, and the single lens magnification
at the time of the planetary deviation determines the separation.
The shape of the deviation indicates a major image perturbation. 
This information is sufficient to specify initial parameters that will
lead to the correct solution.

The other event of modest magnification that is not discussed in this
section is MOA-2007-BLG-197. This event is not included because 
the primary analysis \citep{cassan-moa197} is not yet complete.

For all calculations in this section, we use the second order integration
scheme given in eqs.~\ref{eq-int_rule} and \ref{eq-AB}, with 
$\delta_c = 0.15$. The angular to radial grid-spacing ratio is
adjusted to optimize the calculations based upon the characteristics
of each individual event.


\subsection{OGLE-2005-BLG-71}
\label{sec-ogle-71}


This event \citep{ogle71}
is an $A_{\rm max} \approx 42$ event with a strong central
caustic, cusp approach deviation due to a massive planet with mass ratio of
$q = 7\times 10^{-3}$. The source passes on the side of 
the primary opposite the location of the
planet, and so it approaches the two strong central caustic cusps.
The interval between these cusp approaches is three days, and the
real-time detection of the planetary signal plus good 
weather allowed OGLE to obtain good light curve coverage
over the entire planetary deviation. The OGLE 
coverage is good enough to pin down the basic planetary
parameters, so we include only the OGLE data in our search for
the correct planetary model.

Based on a single lens fit to the OGLE data with the planetary
deviation removed, we set $t_E = 80\,$days, $u_0 = 0.025$, 
and $t_0 = 3481.0\,$days (${\rm JD}-2450000$). This event has no
obvious finite source features, so I fix the source radius crossing time,
$t_\ast = 0$, and search for point-source models. (Finite source effects are
\citep{dong_ogle71} detected in the full analysis of the data, but their
inclusion does not significantly modify the other parameters.)

These cusp approach events are among the easiest planetary events to model
as the fitting code will converge to the correct solution from a large range of
initial condition parameters. So, I set the initial star-planet separation to $d = 0.7$, 
and the planetary mass fraction to either $\epsilon_1 = q/(1+q) = 5\times 10^{-3}$
or $10^{-2}$, and then scan over $\theta = 225^\circ ... 315^\circ$ at
a $1^\circ$ interval. A range of only $90^\circ$ is needed for $\theta$
because we know by inspection the approximate source trajectory.
The fitting code is then started at the parameters that
yield the best initial $\chi^2$ value for each initial $\epsilon_1$ (or $q$) value.
The runs for both initial $\epsilon_1$ values converge to essentially the
same model with $\chi^2 = 280.18$ for 305 data points with
$t_E = 70.96\,$days  $t_0 = 3480.6683\,$days, $u_0 = 0.02352$,
$d = 0.7626$, $\theta = 266.3^\circ$, $\epsilon_1 = 6.79\times 10^{-3}$
(or $q = 6.84\times 10^{-3}$).  There is, of course, also a solution with
$d \approx 1/0.7626 = 1.311$, that can easily be found using this solution
plus the substitution $d \rightarrow 1/d$ as an initial condition.

Because these runs start far from the final solution, we find that it is most
efficient to start at a high Metropolis algorithm temperature, $T$. I then 
reduce $T$ several times during the fit run. For this event, I started at 
$T = 50$, and then dropped it to 5, 0.5, and 0.05 to reach the final solution, which
was reached after 110,559 $\chi^2$ calculations from the 
$\epsilon_1 = 5\times 10^{-3}$ starting point. The run starting from
$\epsilon_1 = 10^{-2}$ required 393,328 $\chi^2$ calculations to approach this
same solution. Despite the large number of $\chi^2$ calculations required,
the entire solution search is fast because point-source calculations were used.
The total calculation took less than 22 cpu minutes on a single cpu of
a 3 GHz Quad-Core Intel Xeon processor (in a MacPro computer purchased
in 2007 running Mac OS 10.5).
The search over the initial condition grid took less than a cpu second.

%

\subsection{OGLE-2003-BLG-235}
\label{sec-moa53}


OGLE-2003-BLG-235Lb was the first definitive exoplanet discovery by
microlensing \citep{bond-moa53}. This event reveals a giant planet
with a mass ratio of $q = 3.9 \times 10^{-3}$ via a caustic crossing
binary lens feature in an event with a modest stellar magnification of
$A_{\rm max} \simeq 7.6$. While it is often the case that the basic
parameters for these lower magnification events can be found by 
inspection, in this case, we have a so-called ``resonant" caustic
with $d \sim 1$, so that the planetary caustic is connected to the
central caustic. In such cases, the caustics are weak, and it can
be difficult to locate the caustic crossings if they are not directly
observed. In the case of OGLE-2003-BLG-235, only the second
caustic crossing was observed, so there is some uncertainty
in the timing of the first caustic crossing.

In order to find candidate solutions for caustic crossing events,
it is most efficient to change variables from $t_0$ and $t_E$ to
the times of the first and second caustic crossings,
$t_{cc1}$ and $t_{cc2} $. This is somewhat similar to, but simpler
than, the scheme of \citet{cassan-meth}.
The calculations for this event were done with a angular to radial
grid spacing ratio of 4 although Figure~\ref{fig-grid_ratio} indicates that
a ratio of 1 would be more efficient.  The mean grid spacing was
0.08 stellar radii.

For OGLE-2003-BLG-235, we fix the following parameters
for the initial condition grid calculation:  
the second caustic crossing
time, $t_{cc2} = 2842.04$ and $u_0 = -0.2216$. The
remaining 5 parameters are allowed to vary over the
following ranges: The planetary mass fraction takes the values
$\epsilon_1 = q/(1+q) = 3.162\times 10^{-4}, 10^{-3}, 3.162\times 10^{-3}, 10^{-2}$.
The separation $d$ takes the values 0.80, 0.84, 0.88, 0.92, 0.96, 
1.04, 1.08, 1.12, 1.16, 1.20, 1.24, and the source trajectory angle ranges over
$\theta = 0^\circ$, ..., $90^\circ$ at $3^\circ$ intervals. The initial condition
grid also includes three source radius crossing times,
$t_\ast = 0.04, 0.07, 0.1\,$days, and 
three first caustic crossing times, $t_{cc1} = 2833.7, 2835.25, 2835.9$.
This gives a total of 12276 initial condition grid points for which I calculate
$\chi^2$. 


Next, the best initial condition for each $t_{cc1}$ value is selected, and used as
an initial condition for $\chi^2$ minimization. This $\chi^2$ minimization is
done with the usual time variables of $t_E$ and $t_0$ instead of
$t_{cc1}$ and $t_{cc2}$. These minimizations are run with an initial
value of $T = 0.5$, which is dropped to $T = 0.05$. Two of the final solutions
match solutions given in \citet{bond-moa53}. The $t_{cc1} = 2835.25\,$days
initial condition leads to the best fit model with 
$\chi^2 = 1641.63$ (for 1535 data points and
1524 degrees of freedom), $d = 1.119$, $\epsilon_1 = 3.9\times 10^{-3}$,
$t_E = 61.78\,$days, and $t_\ast = 0.058\,$days. The $t_{cc1} = 2833.7\,$days
initial condition yields the ``early caustic" model of \citet{bond-moa53}, with
$\chi^2 = 1649.35$, $d = 1.120$, $\epsilon_1 = 6.6\times 10^{-3}$,
$t_E = 59.53\,$days, and $t_\ast = 0.060\,$days.

The modeling run with the $t_{cc1} = 2835.9\,$days initial condition yields
a solution that was not reported in the  \citet{bond-moa53} discovery paper.
This ``late caustic" crossing model has
$\chi^2 = 1646.25$, $d = 1.119$, $\epsilon_1 = 3.4\times 10^{-3}$,
$t_E = 61.3\,$days, and $t_\ast = 0.055\,$days, so it is a somewhat better
model than the ``early caustic" crossing model reported in the paper.
However, these parameters are within 1-$\sigma$ of the best solution
values reported in the discovery paper, so it appears likely this model was
included in the error bar calculations.


These $\chi^2$ minimization runs each included an average of 19,000 $\chi^2$ 
calculations and used about 1.4 cpu hours each. The total number of $\chi^2$
calculations needed to find these three models was slightly over 70,000, and these
calculations were accomplished in 5.24 cpu hours.

\subsection{MOA-2008-BLG-310}
\label{sec-moa310}

\begin{figure}
\plotone{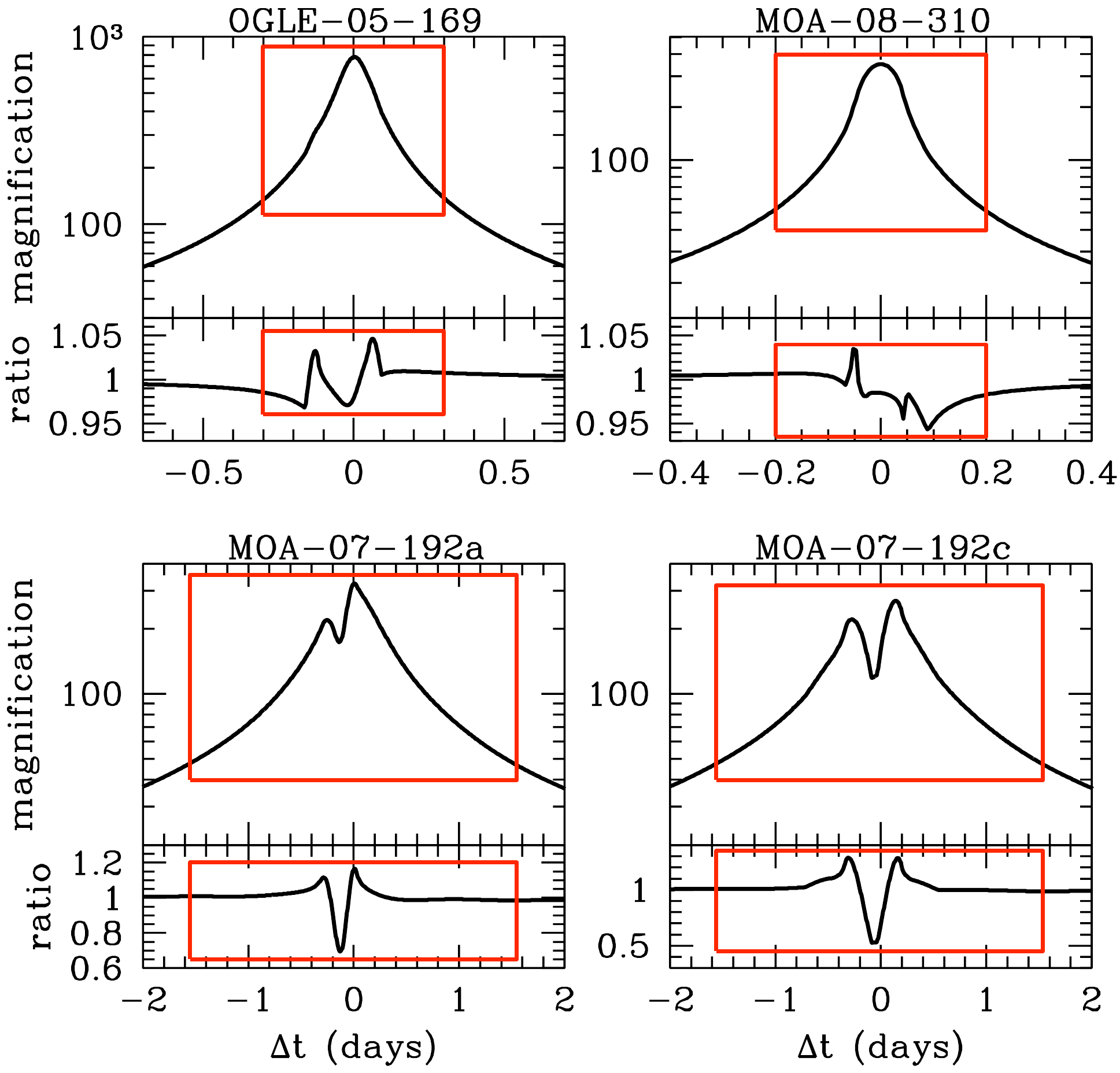}
\caption{Light curves of the high-magnification planetary events used in the tests of
calculation precision are shown. For each light curve, we display the
peak of the light curve on logarithmic scale and the ratio of the planetary
light curve to the single lens light curve with the identical single lens
parameters. As in Figure~\ref{fig-lc_lo}, the red boxes indicate the portions
of the light curves used for the comparisons.
\label{fig-lc_hi}}
\end{figure}
\begin{figure}
\plotone{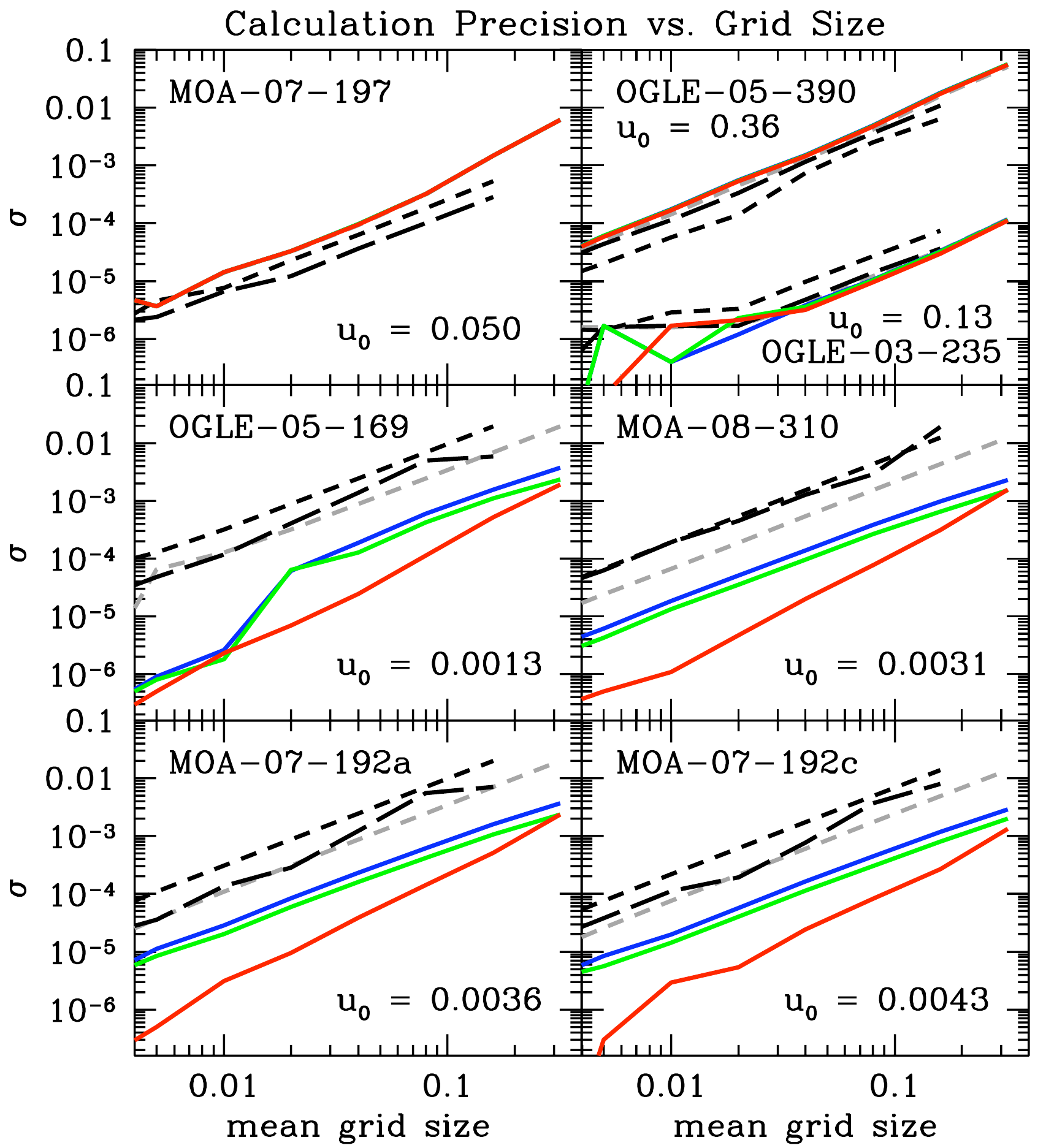}
\caption{The RMS precision, $\sigma$, of the light calculations is shown for our
seven example light curve as a function of the geometric mean grid size in units of the 
source star radius. The blue, green, red, and grey-dashed curves are for
$\delta_c = 0.017$, 0.05, 0.15, and 1.00, respectively. For all these curves,
the angular grid spacing is four times larger than the radial grid spacing
(at the Einstein ring radius). The black short-dashed curve is for $\delta_c = 1.00$, with
equal grid spacing in the angular and radial directions, and the black
long-dashed curve represents a first order integration scheme with no
second order corrections.
\label{fig-compare}}
\end{figure}
\begin{figure}
\plotone{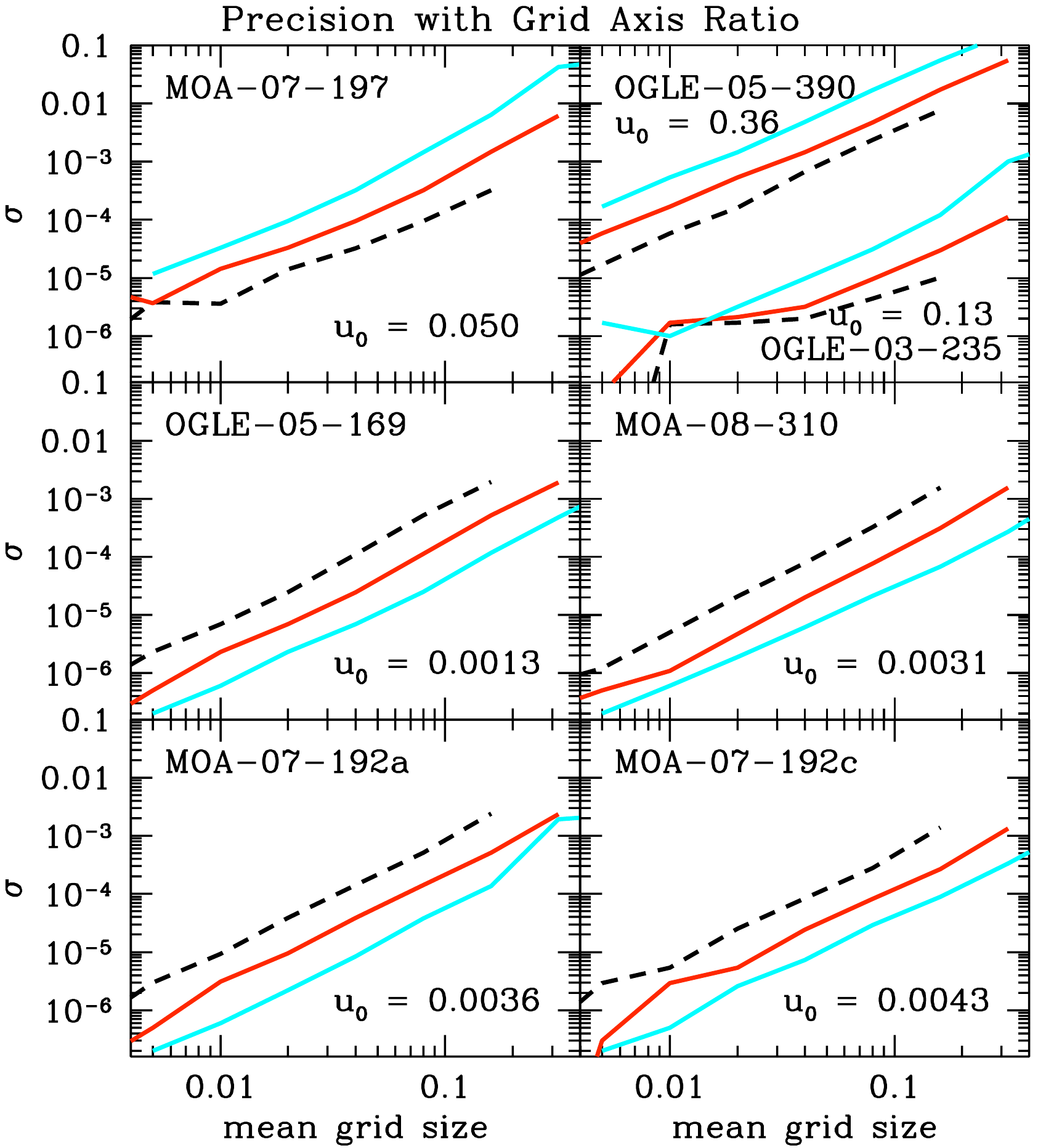}
\caption{The RMS precision, $\sigma$, of the light curve calculations is shown
for the seven example light curves as a function of the geometric mean grid size
(in source star radius units). The cyan, red, and black-dashed curves have 
an angular grid spacing of 16, 4, and 1 times larger than the radial spacing,
respectively. $\delta_c = 0.15$ is used in all cases.
\label{fig-grid_ratio}}
\end{figure}
\begin{figure}
\plotone{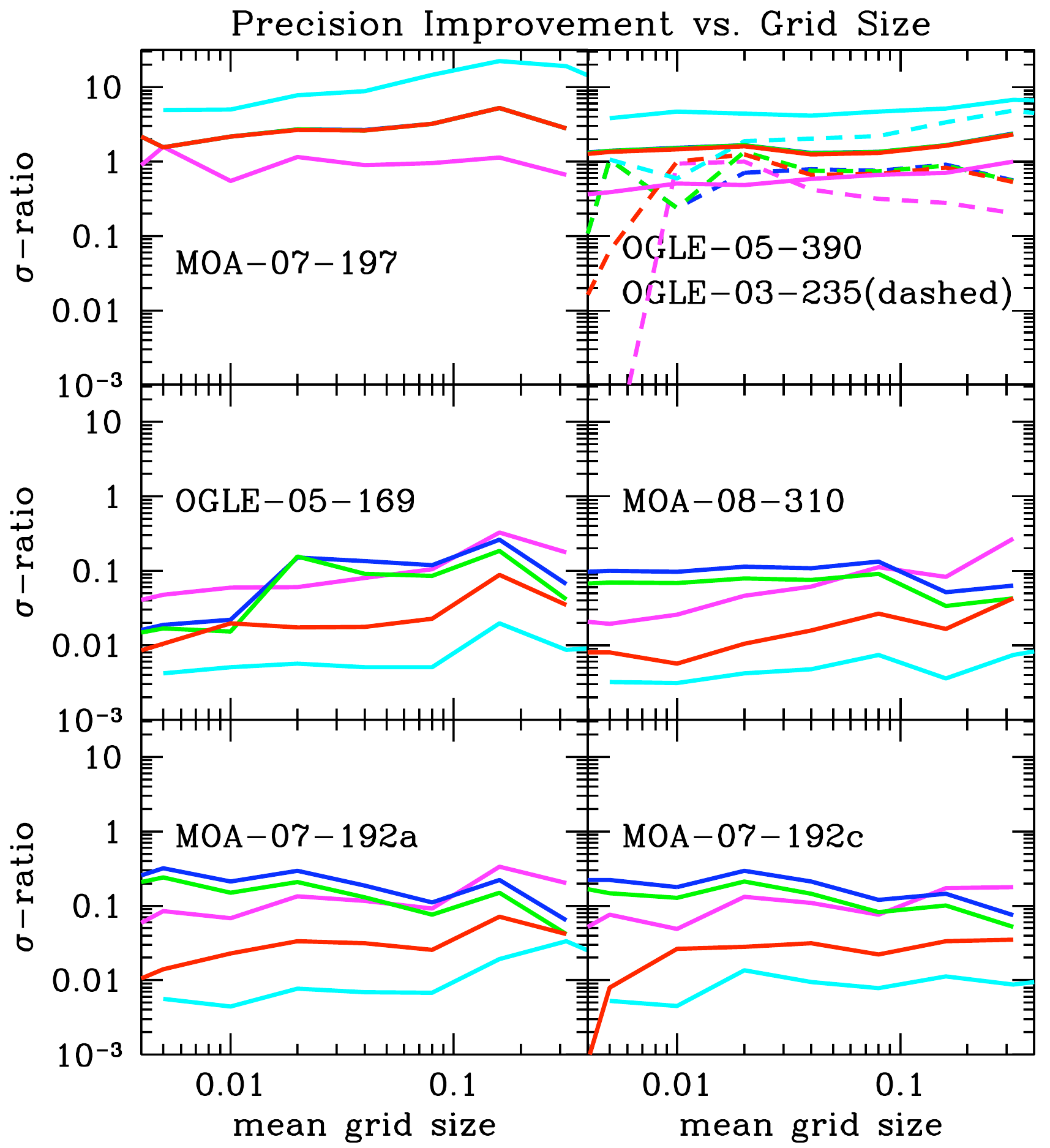}
\caption{The improvement in the RMS precision, $\sigma$, over the first
order integration case (with equal spacing for the
angular and radial grids) is plotted vs.\ the geometric mean grid size.
As in Figures~\ref{fig-compare} and \ref{fig-grid_ratio}, the blue, green,
and red curves have $\delta_c = 0.017$, 0.05, and 0.15, respectively, with
an angular grid spacing 4 times larger than the radial spacing. The magenta and
cyan curves have $\delta_c = 0.15$ and angular grid spacings that are
$1\times$ and $16\times$ larger than the radial grid spacing.
\label{fig-improve}}
\end{figure}

MOA-2008-BLG-310 \citep{moa310} is a high-magnification event
in which the angular radius of the source star is larger than the width
of the central caustic that it crosses. As can be seen from 
Figure~\ref{fig-lc_hi}, the planetary deviation has a maximum 
amplitude of only $\sim 5$\% compared with the corresponding single lens
model. Since this deviation occurs in a region of the light curve
when the magnification due to the stellar lens is changing rapidly,
it is difficult to see the planetary deviation in the raw light curve,
before it is divided by the single lens model.

The light curve deviation due to the planet in Figure~\ref{fig-lc_hi}
does not resemble the light curve deviations for point sources or
sources that are much smaller than the width of the central caustic.
One might worry that this unfamiliar light curve deviation shape
could be a sign that modeling such an event would be difficult.
But, in fact, this is not the case. Events such as this or
MOA-2007-BLG-400 \citep{dong-moa400} turn out to be relatively
easy to model because there are few,
if any, local $\chi^2$ minima besides the global minima for $d < 1$ and 
$d > 1$. As a result, a relatively sparse initial grid is all that is required.

In this case, I have selected an initial grid that is somewhat larger than
is necessary. The single lens parameters, plus the source radius crossing
time, are fixed to the values from the best fit single lens model:
$t_E = 11.022\,$days, $u_0 = 0.002966$,  $t_0 = 4656.399\,$days, and
$t_\ast = 0.05485\,$days. The remaining three binary lens parameters are
scanned over the following ranges:
$\epsilon_1 = 10^{-5}, 10^{-4}, 10^{-3}$; $d = 0.5, 0.6, 0.7, 0.8, 0.9, 1.0$;
and  $\theta = 0^\circ$, ..., $356^\circ$ at $4^\circ$ intervals. Thus, the 
initial condition grid calculations require 1620 $\chi^2$ evaluations.
The best $\chi^2$ values from this grid come from the parameter
sets ($d = 0.9$, $\epsilon_1 = 10^{-4}$,  $\theta = 120^\circ$) and
($d = 0.5$, $\epsilon_1 = 10^{-3}$, $\theta = 116^\circ$). $\chi^2$
minimization runs starting from these initial conditions converge
to the same solution, which corresponds to the $d < 1$ solution
of \citet{moa310}. The parameters of this solution are
$t_E = 10.40\,$days, $t_0 = 4656.3997\,$days, $u_0 = 0.00322$,
$d = 0.921$, $\theta = 112.7^\circ$, $\epsilon_1 = 3.38\times 10^{-4}$,
and $t_\ast = 0.0546\,$days. The wide ($d > 1$) solution is easily 
found from this one. We note that these parameters differ slightly
from those of \citet{moa310} due to slight differences in the
data sets used and a different treatment of the error bars.

These $\chi^2$ minimization runs each required approximately
18,000 $\chi^2$ evaluations, and the combination of the 
initial condition grid search and $\chi^2$ minimizations required
about 7.5 cpu hours using a angular to radial grid spacing ratio of
16 and a mean grid size of 0.16 source radii.




\subsection{OGLE-2005-BLG-169}
\label{sec-ogle169}

OGLE-2005-BLG-169 \citep{ogle169}
is a high-magnification caustic crossing event that
is similar, in some ways, to MOA-2008-BLG-310. However, in this case
the source is much smaller than the width of the central caustic. Like
OGLE-2003-BLG-235, this event does have one well observed caustic
crossing with a caustic crossing time that can be accurately estimated by
inspection of the light curve. But, unlike the case of OGLE-2003-BLG-235,
the stellar magnification peak at $A_{\rm max} =800$ is a much stronger
feature of the light curve than the resolved caustic crossing. As a result,
it is more sensible to use the standard time parameters, $t_E$ and $t_0$
instead of the caustic crossing times, $t_{cc1}$ and $t_{cc2}$.





One difficulty with modeling OGLE-2005-BLG-169 is the incomplete coverage of the
light curve. While the light curve extending from the stellar magnification peak to
$\sim 1.2$ magnitudes below the peak is very densely covered with observations
from the 2.4m MDM telescope every 10 seconds, the rising portion of the light curve
is only observed about once every two hours. As a result, there are several points
in the light curve where the first caustic crossing could have occurred, and 
this complicates the search for a solution.

The caustic crossing observed for this event is quite weak, and this implies that
the planet-star separation must be very close to the Einstein ring. In such
a situation, the shape of the central caustic is a very sensitive function of the 
planet separation and mass fraction, so I use a denser-than-usual 
initial condition grid. The separation, $d$, spans the range from 0.97 to 1.03
at an interval of 0.015, and the planetary mass fraction ranges from 
$\epsilon_1 = 2\times 10^{-5}$ to $\epsilon_1 = 2\times 10^{-4}$ in 
logarithmic intervals of $\sqrt{2}$. For some source trajectories, these
high-magnification, resonant caustic events do not have the usual
$d \leftrightarrow 1/d$ symmetry, so it is prudent to search over both
$d \leq 1$ and $d > 1$. The source trajectory angle spans the
range  $\theta = 0^\circ$, ..., $356^\circ$ with a $4^\circ$ interval. The
observed caustic crossing could, in principle,  be used to determine the 
source radius crossing time, $t_\ast$, but since the first caustic crossing
is unobserved, the angle of the crossing is unknown, we can only be sure that
$t_\ast \leq 0.03\,$days, although a value very much smaller than this would
require an unreasonably shallow crossing angle. I allow $t_\ast$ to range
from $0.01\,$days to $0.03\,$days in the initial condition grid. This parameter
set yields an initial condition grid of 28,350 grid points. The remaining parameters
are fixed at their single-lens fit values: $t_E = 41.63\,$days, 
$t_0 = 3491.8756\,$days, and $u_0 = 0.001256$.

The initial grid search generates four parameter sets that are used for
the subsequent $\chi^2$ minimizations. These include two parameter sets
with $\theta \sim 60^\circ$. These are $\epsilon_1 = 7\times 10^{-5}$, 
$d = 1.015$,  $\theta = 64^\circ$ and $t_\ast = 0.0172\,$days, as well as
$\epsilon_1 = 7\times 10^{-5}$, $d = 0.985$,  $\theta = 56^\circ$,  and
$t_\ast = 0.015\,$days. $\chi^2$ minimization from these initial conditions leads
to two solutions quite similar to the best fit solution of
\citet{ogle169}. The best model has $\epsilon_1 = 8.35\times 10^{-5}$, 
$d = 1.0120$, $\theta = 62.8^\circ$, $t_E = 43.48\,$days, $t_0 = 3491.8756\,$days,
and $t_\ast = 0.0185\,$days with $\chi^2 = 536.29$ for 605 data points
and 588 degrees of freedom. The other model has a slightly worse $\chi^2 = 537.67$
with similar parameters except that  $d = 0.9820$. These correspond to the best
fit model of \citet{ogle169}. The remaining two models have $\theta \approx 95^\circ$,
and they correspond to the secondary local minimal shown in 
Figure 2 of \citet{ogle169}. Note that \citet{ogle169} use a different source trajectory
angle, $\alpha$ that is related to our source trajectory angle by 
$\alpha = 180^\circ - \theta$.

The $\chi^2$ minimization runs for OGLE-2005-BLG-169 averaged about
14,000 $\chi^2$ evaluations and each took about 5.4 cpu hours to complete
using an angular to radial grid spacing ratio of 16 and a mean grid spacing
of 0.16 source star radii. The total number of $\chi^2$ evaluations needed
for OGLE-2005-BLG-169 modeling is 85,400, and this required 32.3 cpu
hours of computing time. This event is the most time consuming of our
example events because of the high-magnification ($A_{\rm max} = 800$)
and the large number of observations at the peak (although the MDM
data are binned to give a sampling interval of 86.4 seconds.)


\subsection{MOA-2007-BLG-192}
\label{sec-moa192}

This event is probably the most challenging of these example events to
model because the sampling of the planetary deviation is quite sparse.
It is a high-magnification event, like OGLE-2005-BLG-169, but slightly
less than half of the peak region is covered with MOA survey observations
with a sampling interval of about 50 minutes. This leads to considerable
uncertainty in the planetary models \citep{bennett-moa192}. In fact,
we cannot be sure if the data indicate a cusp approach or caustic
approach solution. Additionally, the source radius crossing time, $t_\ast$,
is not well constrained. As a result, a relatively large initial condition
grid must be used. The initial parameters are:
$\epsilon_1 = 10^{-5}, 3.16\times 10^{-5}, 10^{-4}, 3.16\times 10^{-4}$; 
$d = 0.5$, 0.6, 0.7, 0.75, 0.8, 0.85, 0.9, 0.95, 0.96, 0.97, 0.98, 0.99, 1.00, 1.01, 
1.02, 1.03, 1.04; $t_\ast = 0.013$,  0.03, 0.047, 0.064, 0.081, 0.099, 
0.116, 0.133, $0.15\,$days; and $\theta = 0^\circ$, ..., $356^\circ$ at 
a $4^\circ$ interval. The remaining parameters are fixed to the values from
a single lens fit with the planetary signal removed:
$t_E = 61.08\,$days, $t_0 = 4245.401\,$days, and $u_0 = 0.00526$.
The total number of $\chi^2$ evaluations in this initial condition grid
is 55,080.



The 12 best results from the grid search were selected for $\chi^2$
minimization. 7 of these $\chi^2$ minimization runs converged to a variant
of the cusp approach solution listed in Table~\ref{tab-modpar}; 4 converged
to a variant of the caustic crossing solution, and the remaining run converged
to another local minimum with a source trajectory angle, $\theta$, 
that differs from the cusp
approach and caustic crossing values by almost $180^\circ$.
This minimum has a $\chi^2$ value larger than the best fit value
by $\Delta\chi^2 > 120$, so it is not considered to be a viable solution.

To get the final set of solutions for this event, we must consider both the cusp
approach and caustic crossing solutions and their $d \leftrightarrow 1/d$
counterparts. Also, this event has a significant microlensing parallax signal
that we do not investigate here, and this introduces other degeneracies
\citep{bennett-moa192}.

\section{Discussion and Conclusions}
\label{sec-conclude}

I have presented a previously unpublished general method for modeling 
multiple mass microlensing events that has been optimized for
high-magnification events. This method has been developed
over a number of years from the first general method for calculating 
binary lens light curves, the image centered ray shooting method 
of \citet{em_planet}. This method is specifically
designed to be computationally efficient for the most demanding high
magnification events, \ie, those with more than two lens masses and/or
orbital motion. There are three aspects to this method: an efficient
method for numerical calculation of the integrals that are needed to
calculation microlensing magnification, an adaptive version of
the Metropolis algorithm to quickly find a $\chi^2$ minimum in
parameter space, and a global search strategy that can find
all the important local $\chi^2$ minima even in parameter
spaces with many dimensions. The computational efficiency
of the first two elements of this method was critical for 
finding the solution for the solution of the only triple-lens 
microlensing system yet to be published \citep{gaudi-ogle109},
as well as the study of the lens masses and orbits that
are consistent with the light curve data for
this event \citep{bennett-ogle109}.

This method has also been successfully tested on all eight published 
single planet microlensing events
\citep{bond-moa53,ogle71,ogle390,ogle169,bennett-moa192,
dong-moa400,moa310,sumi-ogle368}, as well as four planetary
events from the 2009 observing season. For two of these
2009 events, this method found the correct solution before the
planetary signal was completed, using data that spanned less
than 25\% of the planetary deviation (although these events did
have characteristics that were relatively easy to model). The details
of several of these test calculations were presented in 
Section~\ref{sec-global-ex}.

In Section~\ref{sec-lc_calc}, I demonstrated the dramatic improvement
in high-magnification 
light curve calculation precision given by my limb-darkening
optimized integration method and polar coordinate grid with a
much larger angular than radial grid spacing. For the high-magnification
events tested, these features improve the light curve calculation
precision by a factor of $\simgt 100$. Nevertheless, it should still
be possible to make significant additional improvements. One improvement
would be to implement the hexadecapole approximation
\citep{pej_hey,gould-hex} to do finite source calculations,
where the point source approximation does not quite work. This
will reduce the number of lens magnification calculations 
that require the full finite
source integrals. For events without caustic crossings, this
can dramatically improve calculation efficiency
\citep{dong_ogle71}, but for most
caustic crossing events, the improvement is likely to be only a
factor of two or so.

Further modifications to the integration grid scheme I present here
could provide a more dramatic improvement in computational 
efficiency. I have used a 16:1 grid spacing ratio for the most
efficient high-magnification light curve calculations presented here.
However, this is a compromise value. The integration of the
bright images that are generated by the primary lens star could 
probably benefit from a larger ratio, but the smaller images 
(or parts of images) that are directly perturbed by the planet
are done more efficiently with a more modest grid-spacing 
ratio. This compromise could
be avoided by going to an adaptive grid scheme in which the
dimensions of the grid are adjusted to match the distortion of
the images over the entire image plane. Of course, it would be quite
time consuming directly calculate the lens distortions over the entire lens
plane, so this would require a simpler prescription to estimate
the lens distortions. It might be that a direct calculation over a 
coarse grid would work.

Of course, the main goal of this method is to be able to model
complicated microlensing events that have yet to be successfully modeled.
So, the best demonstration of the strength of this method would be the successful
modeling of a number of these events.


\acknowledgments
D.P.B.\ was supported by grants 
AST-0708890 from the NSF and NNX07AL71G from NASA. 

\clearpage


\begin{deluxetable}{lcccccccc}
\tablecaption{Planetary model parameters \label{tab-modpar} }
\tablewidth{0pt}
\tablehead{
\colhead{Event}  & \colhead{$t_E$} & \colhead{$t_1$} &  \colhead{$t_2$} &
       \colhead{$u_0$} &  \colhead{$d$} & \colhead{$\theta$} & \colhead{$q$} &
 \colhead{$t_\ast$} 
}  

\startdata

MOA-07-197  & 64.45 & -11.241 & 26.759 & 0.0500 & 1.156 & 2.247 & $7.64\times 10^{-2}$ & 0.0490 \\
OGLE-03-235  & 61.52 & -13.863 & -5.063 & 0.1327 & 1.120 & 0.764 & $3.94\times 10^{-3}$ & 0.0593 \\
OGLE-05-390  & 11.03 & 8.769 & 11.269 & 0.3589 & 1.610 & 2.756 & $ 7.57\times 10^{-5}$ & 0.282 \\
OGLE-05-169  & 41.72 & -0.301 & 0.299 & 0.00125 & 1.020 & 1.020 & $8.77\times 10^{-5}$ & 0.0184 \\
MOA-08-310  & 10.47 & -0.200 & 0.200 & 0.00314 & 1.094 & 1.961 & $3.51\times 10^{-4}$ & 0.0546  \\ 
MOA-07-192a  & 74.46 & -1.553 & 1.547 & 0.00360 & 1.120 & 4.262 & $1.25\times 10^{-4}$ & 0.0643 \\ 
MOA-07-192c  & 75.05 & -1.562 & 1.538 & 0.00433 & 0.985 & 4.518 & $2.07\times 10^{-4}$ & 0.117  \\ 

\enddata
\tablecomments{ This table shows the model parameters for the 7 events used
for our light curve calculation comparisons. $t_E$ is the Einstein radius  crossing time,
and $t_1$ and $t_2$ are the lower and upper limits of the light curve comparison 
interval, measured with respect to the time of closest approach of the source to the 
lens center-of-mass. $u_0$ is
distance of the closest approach.
$q$ and $d$ are the planet:star mass ratio and separation, and $\theta$ is the angle
between the source trajectory and the planet-star axis.  $t_E$, $t_1$, $t_2$ and $t_\ast$
are all measured in days, while $u_0$ and $d$ are measured in units of
$R_E$, the Einstein radius.}
\end{deluxetable}
\end{document}